\documentclass[manuscript]{aastex}

\shorttitle{Galactic Chemical Evolution and solar $s$-process abundances}
\shortauthors{Bisterzo et al.}

\begin{document}

\title{Galactic Chemical Evolution and solar $s$-process abundances: 
dependence on the $^{13}$C-pocket structure}

\author{S. Bisterzo\altaffilmark{1} and C. Travaglio\altaffilmark{2}}
\affil{INAF - Astrophysical Observatory Turin, Turin, Italy}

\email{bisterzo@to.infn.it; sarabisterzo@gmail.com}

\author{R. Gallino\altaffilmark{2}}
\affil{Department of Physics, University of Turin, Italy}

\author{M. Wiescher} 
\affil{Joint Institute for Nuclear Astrophysics (JINA), Department 
of Physics, University of Notre Dame, IN, USA}

\and

\author{F. K{\"a}ppeler}
\affil{Karlsruhe Institute of Technology, Campus Nord, 
Institut f{\"u}r Kernphysik, Karlsruhe, Germany}

\altaffiltext{1}{Department of Physics, University of Turin, Italy}
\altaffiltext{2}{B2FH Association-c/o Strada Osservatorio 20, 10023 Turin, Italy}

\begin{abstract}
We study the $s$-process abundances (A $\ga$ 90) at the epoch of the solar-system 
formation. AGB yields are computed with an updated neutron capture network  
and updated initial solar abundances.  
We confirm our previous results obtained  with a Galactic Chemical Evolution (GCE) model:
(i) as suggested by the $s$-process spread observed in disk stars and in presolar meteoritic 
SiC grains, a weighted average of $s$-process strengths is needed 
to reproduce the solar $s$-distribution of isotopes with A $>$ 130;
(ii) an additional contribution (of about 25\%) is required in order to represent 
the solar $s$-process abundances of isotopes from A = 90 to 130. 
\\

Furthermore, we investigate the effect of different internal structures of the 
$^{13}$C-pocket, which may affect the efficiency of the $^{13}$C($\alpha$, n)$^{16}$O reaction,
the major neutron source of the $s$-process.
First, keeping the same $^{13}$C profile adopted so far, we modify by a factor of two the
mass involved in the pocket; second, we assume a flat $^{13}$C profile in the pocket, and
we test again the effects of the variation of the mass of the pocket.
\\
We find that GCE $s$-predictions at the epoch of the solar-system 
formation marginally depend on the size and
shape of the $^{13}$C-pocket once a different weighted range of
$^{13}$C-pocket strengths is assumed.
We ascertain that, independently of the internal structure of the
$^{13}$C-pocket, the missing solar-system $s$-process contribution in the range
from A = 90 to 130 remains essentially the same.
\end{abstract}



\section{Introduction}\label{intro}

%
%

The solar-system abundances result from contributions of different
nucleosynthesis processes.
Isotopes heavier than the iron group are produced via neutron captures, the 
$slow$ and the $rapid$ processes ($s$ and $r$). Exceptions 
include the 35 or so\footnote{The original 35 proton-rich nuclei reported by 
\citet{cameron57} and \citet{burbidge57}
 were defined on the basis of being apparently excluded by the $s$- or 
by the $r$-process.} isotopes 
synthesized by the $p$-process.
\\
The origin of the $r$-process is currently attributed to explosive nucleosynthesis in 
massive stars, even if the astrophysical conditions are still not well defined 
(see, e.g., \citealt{thielemann11}). 
The $p$-process is identified with photo-disintegration processes 
(the so-called $\gamma$-process; \citealt{howard91}) 
of 
heavy elements occurring
in core collapse Supernovae of Type II (e.g., \citealt{rauscher02,hayakawa08}). 
Accreting white dwarfs in binary systems 
with masses approaching the Chandrasekhar limit may 
substantially contribute to the production of $p$-process nuclei when 
they explode as Type Ia Supernovae  
(e.g., \citealt{travaglio11}, and references therein).
\\
Concerning the $s$-process, pioneering studies \citep{seeger65,clayton67} have 
demonstrated that the solar $s$-distribution of isotopes from iron to bismuth may be 
reproduced by three components: the $weak$, the $main$ and the $strong$ component. 
The $weak$ component contributes to 
$s$-isotopes up to A $\sim$ 90. It takes place in massive stars during core He and 
convective shell C burning (see, e.g., \citealt{pignatari10}; \citealt*{fris12}).
The $main$ component (A $\sim$ 90 to 208) derives from low mass
asymptotic giant branch (AGB) stars,
during their thermally pulsing (TP) phase 
\citep{arlandini99,goriely00,cristallo09,bisterzo11,lugaro12}. 
The $strong$ component is needed to reproduce 
about half of solar $^{208}$Pb. It originates from
AGB stars of low metallicity (at [Fe/H] $\sim$ $-$1;
\citealt{gallino98,travaglio01,vaneck01}).

After a limited number of TPs, at the quenching of recurrent thermal 
instabilities in the He shell, the convective envelope penetrates in the
outer region of the He-intershell (third-dredge-up, TDU), bringing to the surface 
newly synthesized $^{12}$C and $s$-process isotopes. The whole envelope undergoes
strong mass loss by stellar winds, leaving the degenerate core that eventually
will become a white dwarf.
\\
The major neutron source of the $s$-process in low mass AGB stars ($M$ $<$ 4 $M_\odot$)
is the
$^{13}$C($\alpha$, n)$^{16}$O reaction, which is activated in radiative conditions
during the interpulse phase \citep{straniero95}. 
At the quenching of a TDU, a small amount of hydrogen is assumed 
to penetrate from the envelope into the top layers of the radiative He-rich zone.
At hydrogen re-ignition, a thin $^{13}$C-pocket forms by proton captures on the abundant
$^{12}$C. 
When the temperature
of $\sim$9$\times$10$^7$ K is reached (which corresponds to $kT$ $\sim$ 8 keV), 
$^{13}$C is fully depleted releasing neutrons with a density of about
10$^7$ n/cm$^{-3}$. 
\\
The $^{13}$C in the pocket is of primary origin, directly synthesized in the star 
from the $^{12}$C produced by partial He burning during previous thermal pulse  
and is independent of the initial composition. 
However, the number of free neutrons per iron seed increases with decreasing 
metallicity. Consequently, for a given $^{13}$C-pocket 
strength, the $s$-process flow firstly feeds the $s$-process peak Sr-Y-Zr (at magic neutron
number N = 50), extending up to $^{136}$Ba, then reaches the second $s$-process peak
(Ba-La-Pr-Ce-Nd at N = 82), extending up to $^{204}$Pb-$^{207}$Pb, with a 
progressive increasing ratio of the heavy-$s$ elements (hs) to the light-$s$ elements (ls). 
At even lower metallicity, it mainly feeds $^{208}$Pb (N = 126) at the termination of the 
$s$-process path.
Therefore, the $s$-process in AGB stars is strongly metallicity dependent. 
The complexity of the $s$-process distribution in AGB stars is confirmed by spectroscopic
observations in different stellar populations (planetary nebulae, post-AGB, MS, S, C(N), Ba, CH 
and CEMP-$s$ stars; see, e.g.,
\citealt{smith90,pequignot94,aoki02,abia02,allen06,jonsell06,sharpee07,sterling08,reyniers07,zamora09}).
Furthermore, for a given metallicity a spread in the $s$-process distribution is observed
for each class of stars. A range of $^{13}$C-pocket strengths is needed in order to explain 
this spread (see, e.g., \citealt{busso01,sneden08,kaeppeler11,bisterzo11,lugaro12}).
A similar spread 
is shown by $s$-process isotopic signatures found in presolar meteoritic SiC grains, which
originated in the outflow of AGB stars
(see, e.g., \citealt{lugaro03,clayton04,zinner07}). 

A marginal activation of the $^{22}$Ne($\alpha$, 
n)$^{25}$Mg reaction occurs at the bottom of the advanced convective thermal pulses, where a temperature of 
$T$ $\sim$3$\times$10$^8$ K is reached. 
A short neutron burst is released with a peaked neutron density (up to $N_n$$\sim$10$^{11}$ 
n/cm$^{-3}$), which provides only a few percent of the total neutron exposure, but
affects the abundances of some important isotopes close to the main branchings of the 
$s$-process 
(e.g., $^{85}$Kr and $^{95}$Zr, sensitive to neutron density).

For an extended review on the $s$-process we refer to \citet{kaeppeler11}, and references therein.

Several model uncertainties affect the AGB yields, e.g., the mass loss efficiency, 
the deepness of the TDU, and the number of TPs 
(\citealt{herwig05,straniero06}).
Particularly challenging in AGB modeling is the 
formation of the $^{13}$C-pocket and the physical prescriptions involved.
\citet{iben83} suggested that, as a consequence of the TDU, a sharp discontinuity 
between the H-rich envelope and the He- and C-rich intershell forms, and a partial 
mixing of protons from the envelope into the He-intershell may take place. 
The amount of protons that diffuses into the He-intershell must be small 
to allow the production of $^{13}$C and to limit the further conversion of $^{13}$C to 
$^{14}$N by proton captures. $^{14}$N mainly acts as a neutron poison of the $s$-process via 
the $^{14}$N(n, p)$^{14}$C reaction. 
In the external zone of the pocket (where protons are more abundant), a $^{14}$N-rich 
zone may also form, depending on the proton profile adopted in the AGB model; this 
region plays a minor 
role in the $s$-process nucleosynthesis.
\\
The hydrogen profile 
and, correspondingly, the internal structure 
of the $^{13}$C-pocket may depend 
on the initial mass and metallicity of the AGB, and on the physical mechanisms 
that may compete inside the star itself.
\citet{herwig97} proposed an exponential diffusive overshooting at the borders of all
convective zones. 
FRUITY models by \citet{cristallo09,cristallo11} adopted an 
opacity-induced overshooting at the base of the convective envelope by introducing
in the model an exponentially decaying profile of the convective velocity.
\\
Starting from \citet{langer99}, \citet{herwig03} and \citet{siess04},
rotation was introduced in stellar models to study its impact on the $^{13}$C-pocket 
structure. First studies agree that rotation-induced instabilities reduce the total mass of 
$^{13}$C in the pocket, owing to the $^{14}$N contamination in the $^{13}$C-rich layer, 
compromising the $s$-process efficiency.
\citet{piersanti13} confirm that rotation-induced instabilities modify the mass extension 
of both $^{13}$C and $^{14}$N abundances in the pocket, and their overlap as well. 
Moreover, they suggest that meridional (Eddington-Sweet) circulation may smooth and enlarge the 
$^{13}$C-rich zone of the pocket.
\\
\citet{denissenkov03} demonstrated that a weak turbulence induced by 
gravity waves presents additional alternative for the $^{13}$C-pocket formation.
On the other hand, \citet{busso12} 
revisit the idea that rotation favors mixing 
indirectly, through the maintenance of magnetic dynamo mechanisms, producing the buoyancy 
of toroidal magnetic structures \citep{busso07}.
\\
 Further investigations on rotation, magnetic fields, gravity waves and the interplay between
these several mechanisms will help to shed light on this challenging issue.

The solar system $s$-process distribution is the result of the 
nucleosynthesis of all previous generations of AGB stars 
that have polluted 
the interstellar medium. 
Therefore, a Galactic chemical evolution (GCE) model is required to follow the 
complex evolutionary processes of the Milky Way. 
\citet{travaglio99} showed that AGB yields computed within a weighted average over the
range of $^{13}$C-pockets are needed in the framework of GCE model to reproduce the 
$s$-distribution observed in the solar system.
This is also suggested by 
the spectroscopic $s$-process spread observed in individual stars and the isotopic 
anomalies measured in presolar SiC grains.
\\
GCE calculations succeeded to reproduce the solar abundances of $s$-only isotopes between $^{134}$Ba 
and $^{208}$Pb (\citealt{travaglio01,travaglio04}). 
However, a deficit (of about 25\%) between GCE predictions at the solar epoch and the abundances 
measured in the solar system was found for Sr, Y, Zr, and $s$-process isotopes up to  A = 130, 
including ten $s$-only isotopes from $^{96}$Mo to $^{130}$Xe (see also \citealt{kaeppeler11};
their Fig.~15).
The $weak-s$ process produces isotopes up to Sr ($\sim$10\% to Sr and $\la$5\% to Y and Zr 
isotopes), with a negligible contribution afterwards. 
An additional $r$-fraction of $\sim$10\% was estimated for solar Sr-Y-Zr, evaluating the $r$-contribution
from the prototypical r-II star CS 22892--052\footnote{CS 22892--052 has 
an $r$-process enrichment of $\sim$40 times the solar-scaled composition ([Eu/Fe] = 1.6; 
\citealt{sneden03}).}. 
Summing up, the $s$-, $r$- (and $p$-) 
contributions predicted by current stellar models 
are not sufficient to explain the solar abundances
of light isotopes from $A$ $\sim$ 90 to 130. 
\citet{travaglio04} hypothesized the existence of an additional process 
of unknown origin, called by the authors LEPP (light-element primary process), which must 
supply $\sim$8\% of solar Sr and $\sim$18\% of solar Y and Zr.
Several scenarios have been recently explored, involving 
a primary component in massive stars that comes from the activation of $^{13}$C($\alpha$, 
n)$^{16}$O in the C core when the temperature is low enough to prevent the $^{13}$N($\gamma$, 
p)$^{12}$C reaction from becoming efficient (defined "cold" C-burning component or $cs$-component 
by \citealt{pignatari13}),
and/or a light $r$-process induced by explosive stellar nucleosynthesis,
e.g., in the neutrino-driven winds (\citealt{arcones13} and references therein).

The aim of this work is to investigate 
the influence of one of the major AGB yield uncertainties, the formation of the
$^{13}$C-pocket, on the predicted solar $s$-process distribution, 
from light neutron capture isotopes (to verify the need of LEPP) up to Pb and Bi 
at the end of the $s$-path.
Firstly, in Section~\ref{update} we present updated $s$-percentages 
for isotopes from Kr to Bi with respect to 
\citet{travaglio04}. 
In Section~\ref{pocket}, we test the effect of AGB yields computed
with different choices of the internal structure of the $^{13}$C-pocket
on GCE $s$-predictions at the epoch of the solar-system formation. 
Our results are briefly summarized in Section~\ref{conclusions}.

\section{Updated solar $s$-abundances predicted by GCE}\label{update}

\subsection{The Galactic chemical evolution model}

The general structure of the Galactic chemical evolution (GCE) model adopted 
in this work is the same described by \citet{travaglio04}. 
The GCE model follows the composition of stars, 
stellar remnants, interstellar matter (atomic and molecular gas), and their mutual
interaction, in the three main zones of the Galaxy, halo, thick disk, and thin disk. 
We concentrate on the chemical evolution inside the solar annulus, located 8.5 kpc 
from the Galactic center. The thin disk is divided into independent concentric annuli, 
and we neglect any dependence on Galactocentric radius.
\\
The chemical evolution is regulated by the star formation rate (SFR), initial mass
function (IMF), 
and nucleosynthesis yields from different stellar mass ranges and populations.
The star formation rate has been determined self-consistently as the result of 
aggregation, interacting and interchanging processes of the interstellar gas, 
which may occur spontaneously or stimulated by the presence of other stars.
The stellar contributions from different mass ranges, or from single and 
binary stars are treated separately: we distinguish between single low- 
and intermediate-mass stars ending their life as a He or C-O white dwarf
(0.8 $M_\odot$ $\leq$ $M$ $<$ 8 $M_\odot$), single massive stars exploding
as Type II supernovae (SNeII, 8 $M_\odot$ $\leq$ $M$ $\leq$ 100 $M_\odot$), 
and binary
stars that give rise to Type Ia supernova (SNIa) events.
SFR and IMS are the same adopted by \citet{travaglio04}.

\begin{figure*} 
\includegraphics[angle=0,width=35pc]{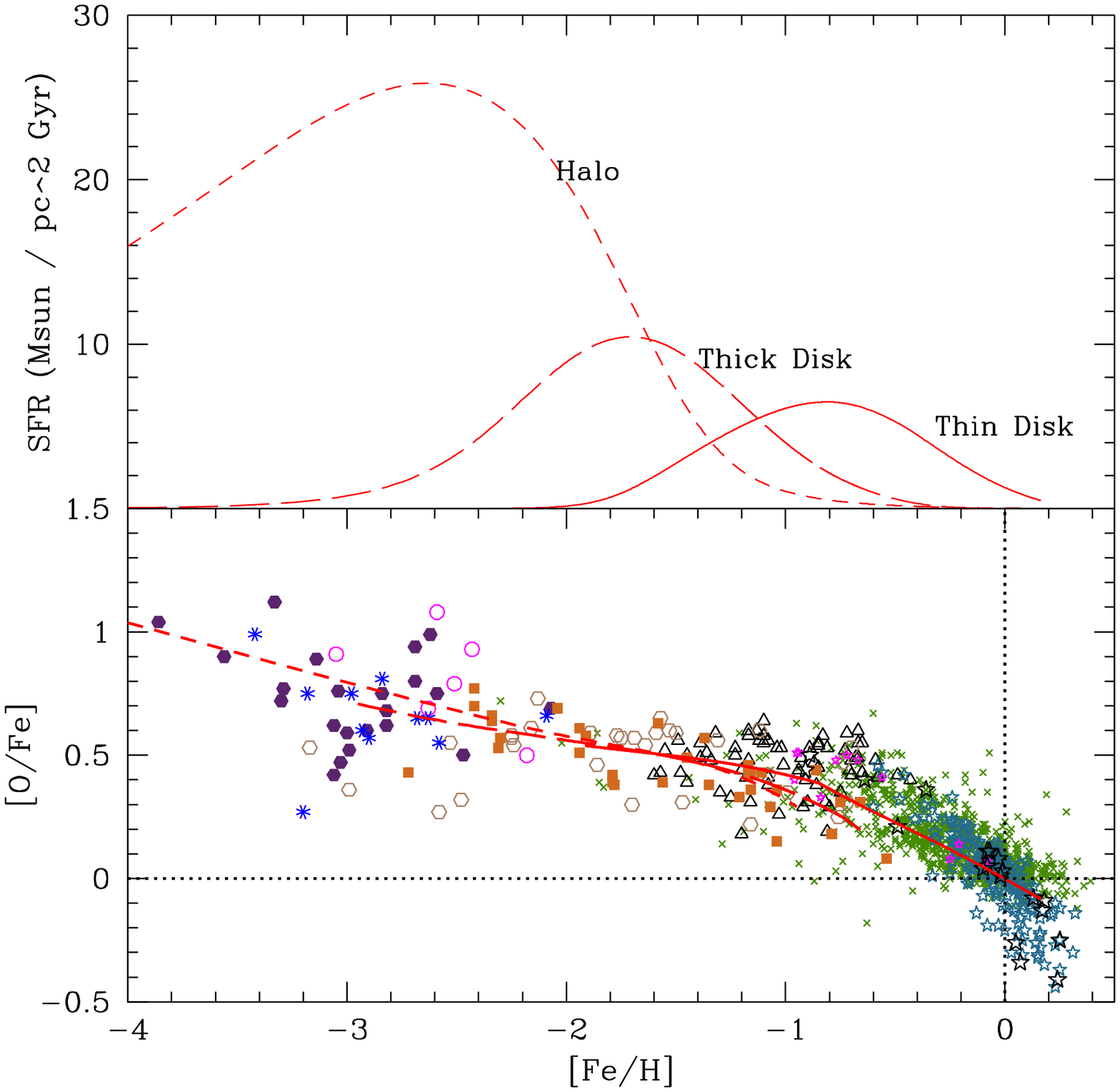}
\caption{\label{SFR} Star formation rate ($top$ $panel$) and [O/Fe] ($bottom$ $panel$) obtained 
with GCE models as a function of [Fe/H].
[O/Fe] predictions are compared with spectroscopic observations by
\citet{spite05} (filled hexagons), 
\citet{lai08} (asterisks), 
\citet{israelian01} (empty circles), 
\citet{melendez06} (empty hexagons), 
\citet{nissen02} (filled squares), 
\citet{ramirez13} (crosses), 
\citet{ramirez12} (empty triangles), 
\citet{mishenina13} (small, middle and big stars for thick, thin,
and unclassified stars, respectively).
{\it See the electronic paper for a color version of this Figure.} }
\end{figure*}

In Fig.~\ref{SFR} ($top$ $panel$), the SFR rate in the three Galactic zones is 
displayed versus [Fe/H]. Galactic halo and thick disk are assumed to form on a 
relatively short time scale: after only $<$0.5 Gyr the metallicity increases up 
to [Fe/H] $\sim$ $-$1.5 (halo phase), to reach [Fe/H] $\sim$ $-$1.0 faster than 
in 2 Gyr. The thin-disk phase in our GCE model starts after $\sim$1 Gyr, at [Fe/H] 
$\sim$ $-$1.5. 
The corresponding [O/Fe] versus [Fe/H] is shown in Fig.~\ref{SFR} ($bottom$ 
$panel$), together with an updated compilation of observational data (see caption). 
Because oxygen is mainly synthesized by 
short-lived SNeII, and under the hypothesis that Fe is mostly produced by long-lived 
SNeIa ($\sim$1/3 by SNeII and $\sim$2/3 by SNeIa), the presence of a ’knee’ in the
trend of [O/Fe] vs [Fe/H] indicates the delayed contribution to iron by SNeIa. 
\\
We refer to \citet{travaglio04} (and references therein), for more details on the 
adopted GCE model.

The Galactic chemical evolution is computed as function of time up to the 
present epoch ($t_{Today}$ = 13.8 Gyr, updated by WMAP; \citealt{spergel03,bennett13}).
Given that the solar system formation occurred 4.6 Gyr ago, the Galactic 
time corresponding to the birth of the solar system is $t_{\odot}$ = 9.2 Gyr,
about 0.7 Gyr later than found by \citet{travaglio04}.
This temporal shift has been achieved by slightly
reducing the Fe contribution by SNeIa. 
\\
Solar abundances have been updated according to \citet{lodders09}, while \citet{travaglio04} 
adopted solar values by \citet{AG89}. 
\\
Stellar yields by \citet{rauscher02} and \citet{travaglio04SNII} 
are used for SNeII and SNeIa, respectively. 
A detailed description of the AGB yields adopted in this work is given in 
Section~\ref{AGByields}.

\subsection{The AGB yields}\label{AGByields}

We have considered a set of five AGB models with low initial mass 
($M$ = 1.3, 1.4, 1.5, 2, and 3 $M_{\odot}$, representing
the AGB mass range 1.3 $\leq$ $M$/$M_\odot$ $<$ 4; hereafter LMS) 
and two AGB models with intermediate initial mass ($M$ = 5 and 7 $M_\odot$, among
4 $\leq$ $M$/$M_\odot$ $<$ 8; hereafter IMS). 
\\
The $s$-process AGB yields have been obtained with a post-process nucleosynthesis
method \citep{gallino98} based on input data (like temperature and density during TPs, 
the mass of the H-shell and He-intershell, the overlapping factor and the residual mass
of the envelope) of full evolutionary FRANEC models 
by \citet{straniero03} for LMS and \citet{straniero00} for IMS.
For details on AGB models we refer to \citet{bisterzo10}. Here we briefly review 
the basic information useful to this work.

As described in Section~\ref{intro},
the total mass of the pocket and the $^{13}$C and $^{14}$N profiles inside the pocket 
itself are not well established, owing to the lack of a definite description of the 
physical mechanisms that allow the diffusion of a few protons into the He intershell.
In our LMS models, 
the amount of $^{13}$C and $^{14}$N in the $^{13}$C-pocket are artificially introduced.
The 
$^{13}$C and $^{14}$N abundances, are treated as free parameters kept constant pulse by pulse.
We adopt a specific shape and size of the $^{13}$C-pocket called case ST (see Section~\ref{pocket}),
which was used to reproduce the solar $main$ component of the $s$-process as the average
of two AGB models with initial mass $M$ = 1.5 and 3.0 $M_\odot$ and half solar metallicity
(see \citealt{arlandini99}).
\\
To interpret the spectroscopic $s$-process spread observed in C and $s$-rich 
stars at a given metallicity, LMS models are computed for a large range of $^{13}$C-pockets,
obtained by multiplying or dividing the $^{13}$C (and $^{14}$N) 
abundances of case ST by different factors, and
leaving the mass of the pocket constant. 
The case ST$\times$2 corresponds to an upper limit, because further proton ingestion leads to 
the formation of $^{14}$N at expenses of $^{13}$C. The minimum $^{13}$C-pocket that significantly 
affects the $s$-distribution depends on the initial iron content of the star,
given that the number of neutrons available per
iron seeds increases with decreasing metallicity. 
As minimum $^{13}$C-pocket, we assume case ST$\times$0.1 at disk metallicities and
case ST$\times$0.03 in the halo (see discussion by \citealt{bisterzo10,bisterzo11}).

LMS models of $M$ = 1.3, 1.4, 1.5 and 2 $M_{\odot}$ have been computed 
following the interpolation formulae given by \citet{straniero03} for $-$1.0 $\leq$ [Fe/H] 
$\leq$ 0.0, and then further extrapolating the stellar parameters from [Fe/H] = +0.2 
down to [Fe/H] = $-$3.6 as described by \citet{bisterzo10}. 
Note that for $M$ = 1.3 $M_{\odot}$ models the conditions for the activation of the TDU 
episodes are never reached for metallicities higher than [Fe/H] $\sim$ $-$0.6 
\citep{straniero03}, and we compute 1.3 $M_{\odot}$ yields starting from [Fe/H] = $-$0.6 
down to $-$3.6.
\\
Starting from \citet{dominguez99} (see also \citealt{boot88}), it was shown that, for a 
given initial mass, the core mass increases with decreasing metallicity and
AGB models with $M$ $\ga$ 3 $M_{\odot}$ and [Fe/H] $\sim$ $-$1.3 are not far from the 
transition zone between LMS and IMS stars. As discussed by \citet{bisterzo10}, we have 
assumed that AGB stars with $M$ = 3 $M_{\odot}$ should behave as IMS for [Fe/H] $\la$ $-$1.6.
Thus, our 3 $M_{\odot}$ models have been have been computed by extrapolating the stellar 
parameters from 3 $M_{\odot}$ models by 
\citet{straniero03} in the metallicity interval between $-$1.6 $\leq$ [Fe/H] $\leq$ +0.2.

We recall that low metallicity IMS stars 
may be affected by an extremely efficient dredge-up called hot TDU 
\citep{herwig04,lau09}, which 
may influence the structure of the star and its evolution. Because nucleosynthesis models 
which include hot TDU are still subject of study, we have computed IMS yields only for
[Fe/H] $\geq$ $-$1.6 \citep{bisterzo10}.

IMS models are based on full evolutionary AGB models by 
\citet{straniero00}. 
Their 5.0 $M_{\odot}$ models at [Fe/H] = 0 and $-$1.3 show comparable characteristics,
e.g., temperature during TPs, TDU and He-intershell masses. 
Thus, we have assumed that the structure of 5.0 $M_{\odot}$ models 
from [Fe/H] = +0.2 down to [Fe/H] = $-$1.6 are barely distinguishable from solar.
Solar 5 and 7 $M_{\odot}$ models described by \citet{straniero00} have
comparable He-intershell mass and temperature at the bottom of the TPs, while the TDU mass
is a factor of $\sim$6 lower in 7 $M_{\odot}$ model. 
Starting from these stellar characteristics, we have extrapolated the 7.0 $M_{\odot}$ models 
in the metallicity range between $-$1.6 $\leq$ [Fe/H] $\leq$ +0.2.
\\
The treatment of mass loss in IMS models is largely uncertain (see, e.g., \citealt{ventura05b}), 
and the number of TPs with TDU experienced by IMS stars may vary by a factor of three or 
more, affecting the structure and nucleosynthesis of the star itself.  
With the efficient mass loss adopted in our IMS models, AGB with $M$ = 5 and 7 $M_\odot$ 
experience 24 TPs followed by TDU \citep{bisterzo10}.

IMS AGB stars play a minor role in the Galactic enrichment of 
$s$-process isotopes, with the exception of a few neutron-rich isotopes as $^{86}$Kr, 
$^{87}$Rb, and $^{96}$Zr \citep{travaglio04}.
Indeed, the mass of the He-intershell is about a factor of ten lower in IMS 
than in LMS.
Thus, the overall amount of material dredged into the envelope of IMS decreases by at 
least one order of magnitude with respect to LMS \citep{straniero00}.  
Moreover, IMS experience hot bottom burning during the interpulse phase: the bottom of 
the convective envelope reaches the top of the H-burning shell and the temperature at the base 
is large enough ($T$ $\ga$ 4$\times$10$^7$ K) to activate the CN cycle, thus converting 
$^{12}$C to $^{14}$N 
(see, e.g., \citealt{karakas03,ventura05}).
Hot bottom burning may even inhibit the formation of the $^{13}$C-pocket in
IMS of low metallicity \citep{goriely04,herwig04}.
As a consequence, the $^{13}$C($\alpha$, n)$^{16}$O neutron source is expected to have 
a small or even negligible effect on neutron capture isotopes heavier than A $\sim$ 90. 
Based on these considerations, new AGB IMS yields have been computed with a negligible $^{13}$C-pocket.  
The introduction of a small $^{13}$C-pocket ($M$($^{13}$C) = 10$^{-7}$ $M_{\odot}$), as assumed in the
old IMS models adopted by \citet{travaglio04}, has negligible effects on solar $s$-process
predictions (see Section~\ref{results}). 
\\
On the other hand, a higher temperature is reached at the bottom of the TPs of IMS ($T$ $\sim$
3.5$\times$10$^8$ K), so that the $^{22}$Ne($\alpha$, n)$^{25}$Mg reaction is efficiently
activated. Because of the high peak neutron density ($N_n$ $\sim$ 10$^{12}$ 
n/cm$^{-3}$), $^{86}$Kr, $^{87}$Rb, and $^{96}$Zr are strongly overproduced, owing to the
branchings at $^{85}$Kr, $^{86}$Rb and $^{95}$Zr. 
The observational evidence of the strong activation of the $^{22}$Ne($\alpha$, n)$^{25}$Mg
reaction (associated to IMS) is the excess of Rb with respect to 
Zr (see, e.g., \citealt{lambert95,tomkin99,abia01}, and the most recent 
\citealt{yong08,garcia09,garcia13,d'orazi13}).
The high [Rb/Zr] observed in some OH/IR AGB stars seems to be incompatible with IMS 
predictions. \citet{karakas12} proposed that delayed superwinds may increase the number of 
TPs, and thus, the [Rb/Fe] and [Rb/Zr] predictions.
A complementary possibility to solve the mismatch between observed and predicted [Rb/Zr] 
in OH/IR AGB stars has been suggested by \citet{garcia09} and \citet{vanraai12}, who 
discussed how model atmospheres of luminous pulsating AGB stars may be incomplete\footnote{E.g., 
circumstellar dust envelope and dust formation, and 3D hydrodynamical simulations are not 
included; NLTE effects are not considered.} and [Rb/Fe] may be overestimated.  
\\
Given the present uncertainties, our 5 and 7 $M_\odot$ models
may experience a larger number of TPs under the hypothesis of a less efficient mass loss.
However, the resulting effect on solar $s$-process predictions is marginal (with the exception of the
neutron-rich isotopes $^{86}$Kr, $^{87}$Rb, and $^{96}$Zr, see Section~\ref{results}), 
because the solar $s$-contribution from LMS stars dominates over IMS. 

In IMS with initial mass of $M$ $\sim$ 6 to 8 $M_\odot$, the temperature at the base of the 
convective envelope further increases ($T$ $\ga$ 1$\times$10$^8$ K), leading to a greater degree
of proton-capture nucleosynthesis that efficiently activates CNO, NeNa and MgAl cycles as well. 
These more massive AGB stars, ending as Ne-O core white dwarfs or possibly less 
massive neutron stars, are called super-AGB stars (SAGB). 
The mass and metallicity limits of SAGB depend on models, 
and the treatment of core carbon ignition and burning and the efficiency of the TDU
is very uncertain (e.g., \citealt{siess10,ventura13}). 
At present, SAGB yields have not been included in our GCE calculations because no information
about heavy $s$-process elements is available in literature \citep{karakas12,doherty14}.

About fourteen hundred AGB models with initial masses $M$ = 1.4, 1.5, 2 $M_\odot$
have been computed in the metallicity range between [Fe/H] = $-$3.6 up to +0.2, for a range
of $^{13}$C-pockets and a total of twenty-eight metallicities. About three hundred AGB models 
with initial masses $M$ = 1.3 $M_\odot$ have been computed between [Fe/H] = $-$3.6 up to $-$0.6, 
for a range of $^{13}$C-pockets and a total of twenty metallicities.
Note that the $s$-process abundances in the interstellar medium at the epoch of the solar system 
formation essentially derive from previous AGB generations starting from [Fe/H] $\ga$
$-$1.6 (see \citealt{travaglio04,serminato09}). Thus, AGB models with a more refined 
grid of twenty metallicities have been computed between [Fe/H] = $-$1.6 and 0.\footnote{Twelve 
metallicities between [Fe/H] = $-$1.6 and $-$0.6 have been run
for 1.3 $M_\odot$ models.}
\\
Similarly, about three hundred 3 $M_\odot$ models have been run between [Fe/H] = $-$1.6 
up to +0.2, for a range of $^{13}$C-pockets and a total of twenty metallicities. 
Forty $M$ = 5 and 7 $M_\odot$ models with a negligible $^{13}$C-pocket have been calculated between 
[Fe/H] = $-$1.6 up to +0.2, for twenty metallicities in total.
\\
Linear interpolations/extrapolations over the whole mass AGB range (1.3 $\le$ 
$M/M_\odot$ $<$ 8, with mass steps of 0.1 $M_\odot$) have been carried out for 
AGB yields not explicitly calculated. 
With respect to \citet{travaglio04}, we have introduced new AGB models of 
initial masses 1.3, 1.4 and 2 $M_{\odot}$.

All AGB yields have been computed with an updated neutron capture  
network according to the online database KADoNiS\footnote{Karlsruhe 
Astrophysical Data Base of Nucleosynthesis in Stars, web site 
http://www.kadonis.org/; version v0.3.}, and more recent measurements:
$^{92,94,96}$Zr \citep{tagliente10,tagliente11,tagliente11a}, 
 $^{186,187,188}$Os \citep{mosconi10},
 the p-only $^{180}$W  \citep{marganiec10},
 as well as 
$^{41}$K and $^{45}$Sc \citep{Heil09pasa},
$^{24,25,26}$Mg \citep{massimi12},
$^{63}$Ni \citep{lederer13},
$^{64,70}$Zn \citep{reifarth12},
 among light isotopes that marginally affect the s-process contribution. 
Note that we employed \citet{mutti05} for $^{80,82,83,84,86}$Kr, 
\citet{patronis04} for $^{136,137}$Cs,
\citet{reifarth03} for $^{148,149}$Pm and \citet{best01} for $^{152,154}$Eu.
\\
Furthermore, AGB initial abundances are scaled to the updated solar values 
by \citet*{lodders09}.

\subsection{Discussion of the results}\label{results}

\begin{figure*} 
\includegraphics[angle=-90,width=40pc]{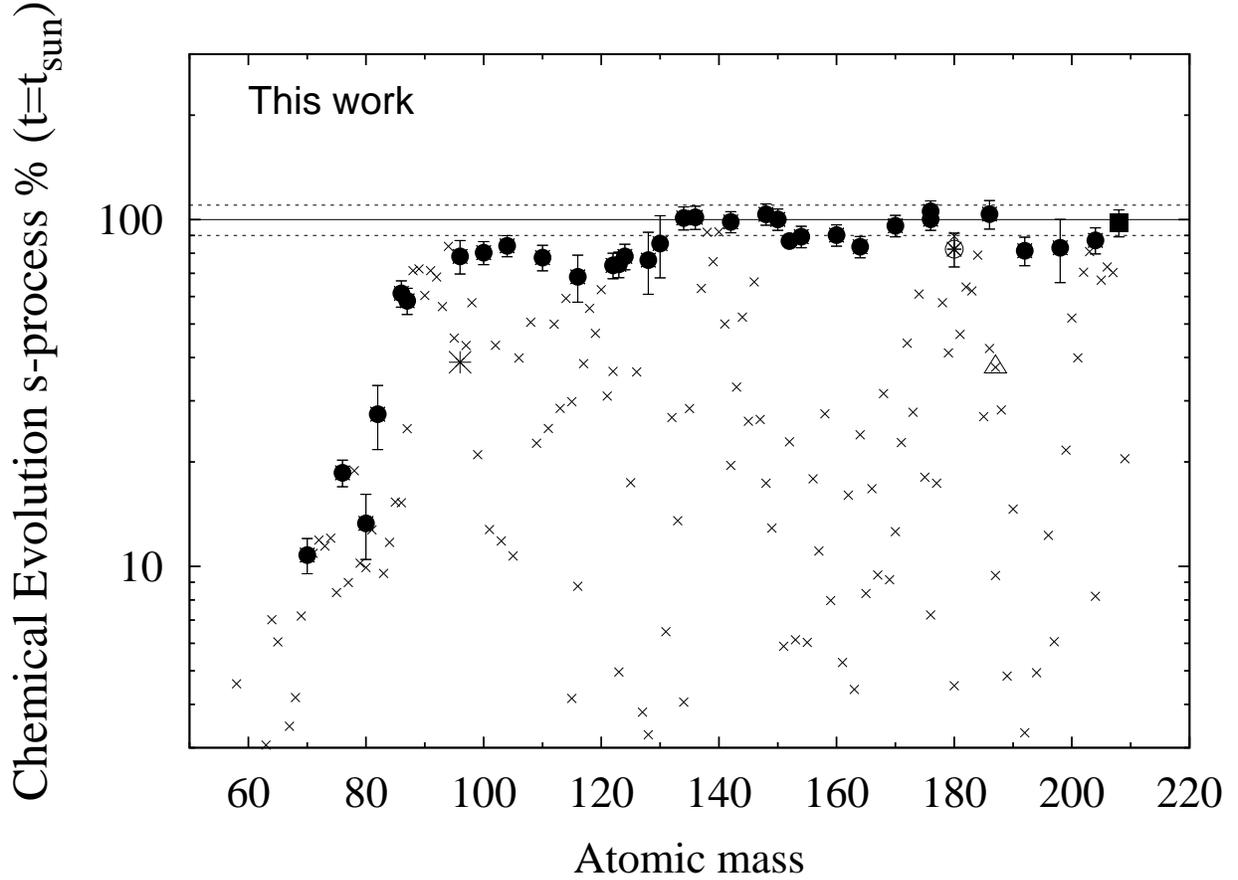}
\caption{\label{Fig1} Reproduction of the solar $s$-process abundances (in \%)  
obtained at the epoch of the solar system formation with GCE model. 
We display the updated solar $s$-process predictions presented in this work.  
The $s$-only isotopes are indicated by $filled$ $circles$. 
Different symbols have been used for 
$^{180}$Ta ($open$ $circle$), $^{187}$Os ($open$ $triangle$), $^{208}$Pb ($filled$ $square$) 
and $^{96}$Zr ($big$ $asterisk$), see text.
Errors account for solar abundance uncertainties.
A full list of the solar $s$-process abundances is given in Table~\ref{Tab1}.}.
\end{figure*}

The $s$-process GCE predictions at the epoch
of the solar system formation are displayed in Fig.~\ref{Fig1} as percentages of the 
abundances by \citet{lodders09}.
The $s$-only isotopes are represented by solid circles. 
Different symbols have been used for $^{96}$Zr (big asterisk),
$^{180}$Ta (empty circle), which also receives 
contributions from the $p$ process and from neutrino-nucleus interactions in massive stars, 
and $^{187}$Os (empty triangle), which is affected by the long-lived 
decay of $^{187}$Re (4.1$\times$10$^{10}$ yr).
We distinguish $^{208}$Pb, which is produced at $\sim$50\% by LMS of low metallicity,
with a big filled square. 
\\
In the framework of GCE we have considered a 
weighted average among the various $^{13}$C-pocket strengths (see Section~\ref{intro}).
Observations
of $s$-process rich stars at disk metallicity 
(see, e.g., 
\citealt{kaeppeler11}, their Fig.~12) suggest that most of them lie in the range 
between case ST$\times$1.5 down to ST/1.5. Very few stars can be interpreted by case 
ST$\times$2. Actually, for $^{13}$C-pockets below case ST/6 the $s$-enhancement 
becomes negligible. We exclude case ST$\times$2 from the average of the 
$^{13}$C-pockets, and we give a weight of $\sim$25\% to case ST$\times$1.3.
The unbranched $s$-only $^{150}$Sm can be taken as reference isotope for the whole
$s$-process distribution, because 
its solar abundance is well defined (5\% uncertainty as a rare earth element, "REE") 
and the neutron capture cross section is given at 1\%.

A complete list of $s$-process percentages of all isotopes and elements 
from Kr to Bi is given in Table~\ref{Tab1},
where updated GCE results are compared with previous GCE predictions by \citet{travaglio04}.
The theoretical $s$-process predictions by \citet{travaglio04} 
were normalized to solar abundances by \citet{AG89}.
The $s$-percentages of LMS, IMS and the total $s$-contribution (LMS + IMS) 
are reported for GCE results by \citet{travaglio04} (columns~3, ~4 and 
~5) and this work (columns~6, ~7 and ~8). Values in column~8 correspond to those displayed 
in Fig.~\ref{Fig1}. 
In comparison to GCE results, we list in column~2 the $main-s$ process contributions 
by \citet{bisterzo11}, obtained as in \citet{arlandini99} by averaging between AGB models 
of $M$ = 1.5 and 3 $M_\odot$ and half solar metallicity.

The variations between GCE results by \citet{travaglio04} and those computed in
this work (see Table~\ref{Tab1}) are partly due to the updated nuclear reaction network 
and new solar abundances. 

\begin{figure*} 
\includegraphics[angle=-90,width=40pc]{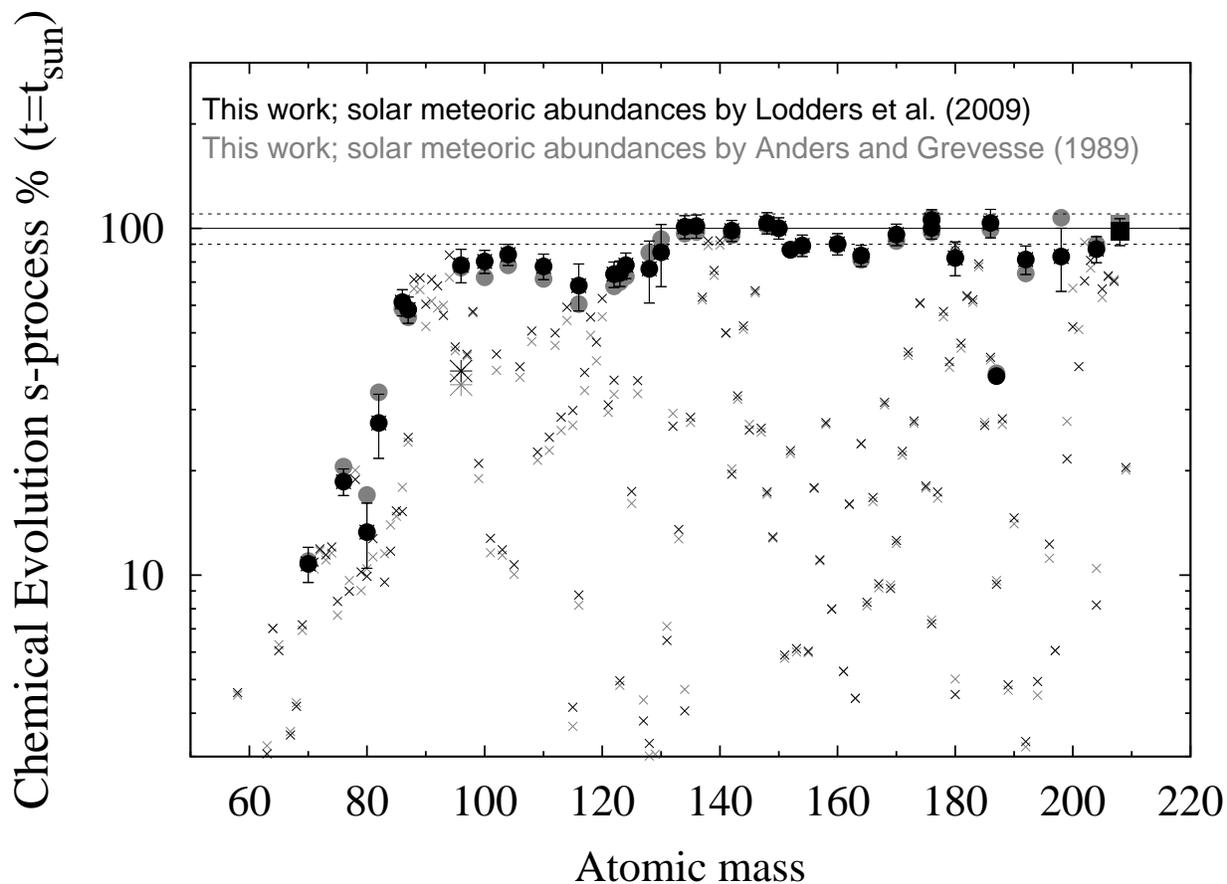}
\caption{\label{Fig2} 
We compare updated solar $s$-process contribution (normalized to solar 
abundances by \citealt{lodders09}; $black$ $symbols$) with the same results
normalized to solar abundances by \citet{AG89} ($grey$ $symbols$).
The $s$-only isotopes are indicated by solid circles, $^{208}$Pb by solid square  
and $^{96}$Zr by big asterisk.}
\end{figure*}

The solar abundance distribution is essentially based on CI carbonaceous chondrite 
measurements in terrestrial laboratories with increasingly accurate experimental 
methods. Among the exceptions are volatile elements as CNO, which are evaluated from 
photospheric determinations, or noble gases as Ne, evaluated from solar corona winds.
Kr and Xe solar abundances are derived 
theoretically from neutron-capture element systematics \citep{palme93,reifarth02}.
Since 1989, C, N and O abundances have been steadily revised downward thanks to 
progress in atmospheric models (see, e.g., \citealt{asplund05}), partially affecting
the isotopic mass fraction of heavier isotopes\footnote{Note that about half of solar 
Z abundances (where Z is the sum of all elements heavier than helium,
called "metals" in stars) comes from oxygen, followed by carbon, neon, nitrogen 
(silicon, magnesium and iron as well).}.
Moreover, the photospheric abundances observed today are different from those existing
at the beginning of the solar system (4.6 Gyr ago), because of the element settling 
from the solar photosphere into the Sun's interior. 
\citet{lodders09} found that original protosolar values of
elements heavier than He (Z) were $\sim$13\% larger than observed today (whereas 
He increases by about $\sim$15\%).
\\
In Fig.~\ref{Fig2}, we show updated GCE results normalized to solar 
abundances by \citealt{lodders09} ($black$ $symbols$) 
and \citet{AG89} ($grey$ $symbols$; as made by \citealt{travaglio04}).
Noteworthy solar abundance variations\footnote{Variations refer to values normalized to 
the number of silicon atoms $N$(Si) = 10$^6$.} ($>$10\%) 
are CNO ($\sim-$30\%), P ($\sim-$20\%), S ($-$25\%), Cl (+37\%),
Kr (+24\%),
Nb (+13\%), I (+22\%), Xe (+16\%),
and Hg (+35\%).

Additional differences between the GCE results shown in 
Table~\ref{Tab1}
derive from 
experimental cross section measurements and 
theoretical nuclear improvements provided in the last decade. 
\\
New GCE calculations include the recent theoretical evaluation of the $^{22}$Ne($\alpha$, 
n)$^{25}$Mg rate by \citet{longland12}, which is based on the direct experimental measurement 
by \citet{jaeger01}.
At AGB temperatures ($T_8$ $\sim$ 2.5 to 3.5) the $^{22}$Ne($\alpha$, n)$^{25}$Mg rate 
by \citet{longland12} is very close to the value by \citet{jaeger01}, and it is
about a factor of two lower than the rate we used so far: 
the lowest limit by \citet{kaeppeler94}\footnote{The lowest limit by \citet{kaeppeler94} has been 
evaluated by neglecting the resonance contribution at 633 keV. Note that, at AGB temperatures, the  
$^{22}$Ne($\alpha$, $\gamma$)$^{26}$Mg rates recommended by \citet{jaeger01} and \citet{longland12} 
are in agreement with the lower limit evaluated by \citet{kaeppeler94}.}. 
The $^{22}$Ne($\alpha$, n)$^{25}$Mg neutron source produces variations of the $s$-isotopes close 
to the branching points: 
updated GCE calculations have reduced the $s$-contribution to few light isotopes,
$^{86}$Kr, $^{85,87}$Rb, $^{96}$Zr (and the $s$-only $^{96}$Mo), with a 
complementary increase of $^{86,87}$Sr, 
owing to the branches at $^{85}$Kr, $^{86}$Rb and $^{95}$Zr. 
Minor effects ($<$10\%) are seen for heavier isotopes  
(see \citealt{bisterzo13}, Fig.~1, bottom panel). A discussion of the major branching points
of the $s$-process and the effect of the new $^{22}$Ne($\alpha$, n)$^{25}$Mg rate will be 
given in Bisterzo et al., in preparation.
\\
Particularly large was the total $s$-production of $^{96}$Zr obtained in 2004 
($\sim$80\%).
Updated GCE results predict that about half of solar $^{96}$Zr comes from
AGB stars ($\sim$3\% from IMS and $\sim$36\% from LMS; see Table~\ref{Tab1}).  
\citet{travaglio11} estimated an additional non-negligible $p$-process contribution 
to $^{96}$Zr by SNeIa (up to 30\%). 
\\
However, we remind that $^{96}$Zr is strongly sensitive to the 
$^{95}$Zr(n, $\gamma$)$^{96}$Zr rate, which is
largely uncertain being evaluated only theoretically. In this regard, we adopt a rate
roughly half the value recommended by \citet{bao00}, close to that by \citet{toukan90},
in agreement with Zr isotopic measurements in presolar SiC grains \citep{lugaro03}.
Recently, \citet{lugaro14} provide a lower estimation of the $^{95}$Zr(n, $\gamma$)$^{96}$Zr rate:
at $kT$ $\sim$ 23 keV, their value is roughly three times lower than that by 
\citet{bao00} and about half of that by \citet{toukan90}. This results in a decreased 
GCE $s$-contribution to solar $^{96}$Zr by $\sim$10\%.
\\
Moreover, $^{96}$Zr depends on the number of 
TPs experienced by the adopted AGB models. 
Our LMS models with $M$ = 1.5, 2 and 3 $M_\odot$ experience several TPs with 
TDU, from 20 to 35 TPs
depending on the mass and metallicity (\citealt{bisterzo10}, their Table~2), 
reaching temperatures high enough to partly
activate the $^{22}$Ne($\alpha$, n)$^{25}$Mg
neutron source and to open the branch at $^{96}$Zr. 
A stronger mass loss would partly reduce 
the $s$ contribution to $^{96}$Zr from LMS AGB stars.
\\
On the other side, our $M$ = 5 and 7 $M_\odot$ models experience 24 TPs. 
Given the large uncertainty affecting the treatment of mass loss in IMS 
(Section~\ref{AGByields}), the number of TPs may increase by a factor
of three or more (in particular for $M$ = 7 $M_\odot$ models), leading to
a greater overall amount of material dredged up.
Owing to the dominant $s$-contribution from LMS stars over IMS, the general effect 
on solar GCE prediction would be negligible even by increasing the mass of the 
TDU in both $M$ = 5 and 7 $M_\odot$ models several times.
Exceptions are $^{86}$Kr, $^{87}$Rb, and $^{96}$Zr, which receive a non-negligible
contribution from IMS stars: by increasing the TDU mass of IMS by a factor of six, 
the solar prediction to $^{96}$Zr increases 
from $\sim$40\% up to $\sim$60\%; similar variations affect $^{86}$Kr ($\sim$+20\%), 
larger effects are show by $^{87}$Rb (from $\sim$30\% up to $\sim$70\%).
\\
As discussed in Section~\ref{AGByields}, new IMS yields are computed with a negligible 
$^{13}$C-pocket; instead, \citet{travaglio04} assumed that a $^{13}$C-pocket
with $M$($^{13}$C) = 10$^{-7}$ $M_{\odot}$ could form in IMS, yielding an increase of
$\sim$10\% to solar $^{96}$Zr, $\sim$3\% to $^{86}$Kr and $\sim$6\% to $^{87}$Rb. 
Solar GCE predictions of isotopes with $A$ $>$ 100 show marginal variations,
because the $^{13}$C-pocket contribution from IMS is almost negligible (up to a 
few percent) when compared to that of LMS.

New neutron capture cross section measurements further modify the $s$-process
distribution.
\citet{travaglio04} adopted the compilation by \citet{bao00}, while
we refer to KADoNiS
(as well as few additional measurements listed above). 
Important variations come from the new $^{139}$La(n, $\gamma$)$^{140}$La rate 
by \citet{winckler06}, the $^{180}$Ta$^m$(n, $\gamma$)$^{181}$Ta rate 
by \citet{wisshak04} (including the revised stellar enhancement factor, SEF, by KADoNiS), 
and the improved treatment of the branch at $^{176}$Lu, which modifies the 
$^{176}$Lu/$^{176}$Hf ratio \citep{heil08}.
The solar $s$-contribution to La, often used as a reference element to disentangle 
between $s$ and $r$ process enrichment in the atmosphere of peculiar stars, increases 
by $\sim$+10\% (from 63\% to 76\%, see Table~\ref{Tab1}). 
The solar $s$-production of $^{180}$Ta$^m$ increases from $\sim$50\% to $\sim$80\%, 
substantially reducing the contribution expected from the $p$-process.

Discrepancies of $\sim$10\% for  
$^{154}$Gd and $^{160}$Dy, 
and $\sim$20\% for 
$^{192}$Pt (see Section~\ref{Pt}) suggest that large uncertainties affect 
the neutron capture cross sections at stellar energy (e.g., SEF evaluation) for 
isotopes with $A$ $>$ 150.
Note that $^{160}$Dy may receive a small contribution from $p$-process \citep{travaglio11}.
Despite all Pb isotopes have well determined neutron capture cross sections
(see KADoNiS), the contribution to the $s$-only $^{204}$Pb is affected by the branch 
at $^{204}$Tl, which produces variations of $\sim$10\%.

Finally, as recalled above, the new GCE predictions adopt an extended 
range of AGB yields, including $M$ = 1.3, 1.4 and 2 $M_{\odot}$ models.
In \citet{travaglio04}, only $M$ = 1.5 and 3 $M_{\odot}$ models among LMS were used for
GCE calculations.
\\
AGB stars with initial mass of $M$ = 1.3 and 1.4 $M_{\odot}$ experience five and ten TPs followed by TDU, 
respectively \citep{bisterzo10}. This results in an additional $s$-process contribution of
about 5\% and 10\% at the epoch of the solar system, without modifying the $s$-distribution.
Note that for metallicities higher than [Fe/H] $\sim$ $-$0.6, $M$ = 1.3 $M_{\odot}$ models 
do not contribute to the chemical evolution of the Galaxy because 
the conditions for the activation of the TDU episodes are never reached in our models
(see \citealt{straniero03}).
\\
The addition of $M$ = 2 $M_{\odot}$ yields in GCE computations provides an increase of about 10\%
of the predicted solar $^{208}$Pb, and smaller variations to isotopes with A $\la$ 140 (lower 
than 5\%), with the exception of solar $^{96}$Zr (which is 10\% lower).
AGB stars with $M$ = 1.5 and 2 $M_{\odot}$ predict comparable $s$-process abundances because
both models reach similar temperatures during convective TPs and the larger overall amount of material 
dredged up by the 2 $M_{\odot}$ model (which experiences about six more TDUs than the 1.5 $M_{\odot}$ model) 
is diluted with a more extended envelope \citep{bisterzo10}. 
Thus, in both 1.5 and 2 $M_{\odot}$ models the marginal activation of the $^{22}$Ne($\alpha$,
n)$^{25}$Mg 
produces a smaller $s$-process contribution to the first $s$-peak (including $^{96}$Zr) than
in 3 $M_{\odot}$ 
models, while on average $^{208}$Pb is favored, being the $^{13}$C($\alpha$, n)$^{16}$O reaction 
the main source of neutrons. 
Instead, AGB stars with initial mass of $M$ = 3 $M_{\odot}$ and [Fe/H] $\sim$ $-$1 reach 
higher temperatures at the bottom of the TPs ($T_8$ = 3.5), feeding more efficiently 
the first %
and second $s$-peaks than $^{208}$Pb.
As consequence, linear interpolations between $M$ = 1.5 and 3 $M_{\odot}$ yields, as previously made by
\citet{travaglio04}, induced up to 10\% differences in $^{96}$Zr and $^{208}$Pb solar predictions.

Updated GCE calculations plausibly interpret within the nuclear and solar uncertainties 
all $s$-only isotopes with $A$ $>$ 130.
\\
The understanding of the origin of light $s$-process isotopes with $A$ $\leq$ 130
remains enigmatic: we confirm the missing $\sim$25\% predicted solar $s$-contribution to isotopes 
from $^{86}$Sr to $^{130}$Xe
(including the $s$-only $^{96}$Mo, 
$^{100}$Ru, $^{104}$Pd, $^{110}$Cd, $^{116}$Sn, $^{122,123,124}$Te and $^{128,130}$Xe)
 firstly found by \citet{travaglio04}.
 Few of them may receive an additional small contribution from the $p$-process
(e.g., 2\% to $^{96}$Mo, 5\% to $^{110}$Cd, 3\% to $^{122}$Te and 6\% to $^{128}$Xe; 
\citealt{travaglio11}).

In Section~\ref{pocket}, we explore the effect of the $^{13}$C-pocket uncertainty in LMS on 
 GCE predictions at the epoch of the solar-system formation.

\subsection{New neutron capture rates for Pt and $^{192}$Ir isotopes}\label{Pt}

The solar $s$-only $^{192}$Pt predicted by GCE model is about 20\% lower than 
measured in the solar system.
So far, we considered this value plausible within the known uncertainties: the KADoNiS 
recommended $^{192}$Pt(n, $\gamma$)$^{193}$Pt rate has 20\% of uncertainty 
(590 $\pm$ 120 mbarn at 30 keV; \citealt{bao00}), 
the neutron capture cross sections of $^{191}$Os and $^{192}$Ir are evaluated theoretically at 
22\%, and the solar Pt abundance is know with 8\% of uncertainty \citep{lodders09}.

Recently, \citet{koehler13} measured the neutron capture cross sections of Pt isotopes 
with much improved accuracy, and used their experimental results to provide
a new theoretical estimation of the $^{192}$Ir(n, $\gamma$) rate.
The new Maxwellian-Averaged Cross Sections (MACS) of $^{192}$Pt(n, $\gamma$)$^{193}$Pt 
(483 $\pm$ 20 mbarn at 30 keV; 4\% uncertainty) 
is $\sim$20\% lower than that recommended by KADoNiS, suggesting an increased  
$^{192}$Pt prediction ($\sim$90\%), in better agreement with the solar value.
On the other side, the theoretical $^{192}$Ir(n, $\gamma$) rate estimated by \citet{koehler13} 
is much larger (about +50\%) than that by KADoNiS (2080 $\pm$ 450 mbarn at 30 keV), reducing 
again the $s$-contribution to $^{192}$Pt. 
\\
The new GCE prediction of solar $^{192}$Pt obtained by including neutron capture rates of 
Pt isotopes and $^{192}$Ir by \citet{koehler13} is $\sim$78\%. This value may increase
up to $\sim$85\% by considering the uncertainty of the theoretical $^{192}$Ir(n, $\gamma$) 
rate ($\pm$22\%), and up to 93\% by adopting a 2$\sigma$ uncertainty.
\\
More detailed analyses on $^{192}$Ir MACS would help to improve the understanding of
this branching point and to provide a more accurate solar $^{192}$Pt estimation (e.g., \citealt{rauscher12}).

\section{Effect of the $^{13}$C-pocket uncertainty in LMS on GCE predictions at the epoch of the 
solar-system formation}\label{pocket}

As discussed in Section~\ref{intro}, the problem of the formation of the $^{13}$C-pocket 
is still unsolved. The tests on the $^{13}$C-pocket discussed in this work allow us to 
explore the impact of different shapes and sizes of the adopted $^{13}$C-pocket on the $s$-process yields.

The internal structure of the $^{13}$C-pocket adopted so far is specified in Table~\ref{Tab2} 
(first group of data): it is a three-zone $^{13}$C-pocket (zones I-II-III roughly correspond 
to those described by \citealt{gallino98}, their Fig.~1), each one has defined $X$($^{13}$C) and $X$($^{14}$N) 
abundances, and a total $^{13}$C mass of $\sim$5$\times$10$^{-6}$ $M_\odot$ ($\sim$2$\times$10$^{-7}$ 
$M_\odot$ of $^{14}$N).
The total mass of the three-zone pocket is about the twentieth part of a typical convective 
TP in LMS ($M$(pocket) $\sim$ 0.001 $M_\odot$).
This corresponds to the so-called case ST. 
As anticipated in Section~\ref{AGByields}, case ST is calibrated to reproduce 
the solar $main$ component \citep{arlandini99}.
\\
Note that case ST differs from a H profile that exponentially decreases 
starting from the envelope value $X$(H) = 0.7. 
In such a case an outer $^{14}$N-rich zone, also called $^{14}$N-pocket,
forms in the outer layers of the He-intershell 
\citep{goriely00,cristallo09,karakas10,lugaro12}.

To evaluate the impact of different $^{13}$C-pocket structures on GCE predictions at the solar
epoch, we made a few tests in which we introduced different $X$($^{13}$C) profiles 
and total masses in the pocket, both considered as free parameters in LMS.
We remind that the effect of the $^{13}$C-pocket in IMS is negligible for GCE predictions at $t_\odot$ = 9.2 Gyr (see
Section~\ref{update}).
\\
Starting from the three-zone $^{13}$C profile used so far, we exclude the two external 
$^{13}$C-rich zones of the pocket (leaving zone II only) in order to obtain flat 
$^{13}$C and $^{14}$N profiles.
We still adopt a range of $^{13}$C-pocket strengths by multiplying the abundances of $^{13}$C and 
$^{14}$N from 0.1 to 2 times the values of zone II.
\\
In addition, we test the effects of the variation of the mass of the pocket $M$(pocket), 
both on the three-zone $^{13}$C-pocket and zone-II $^{13}$C-pocket. 
\\
The internal structure of the $^{13}$C-pocket adopted in the different tests is specified
in Table~\ref{Tab2}.

\begin{figure*} 
\includegraphics[angle=-90,width=40pc]{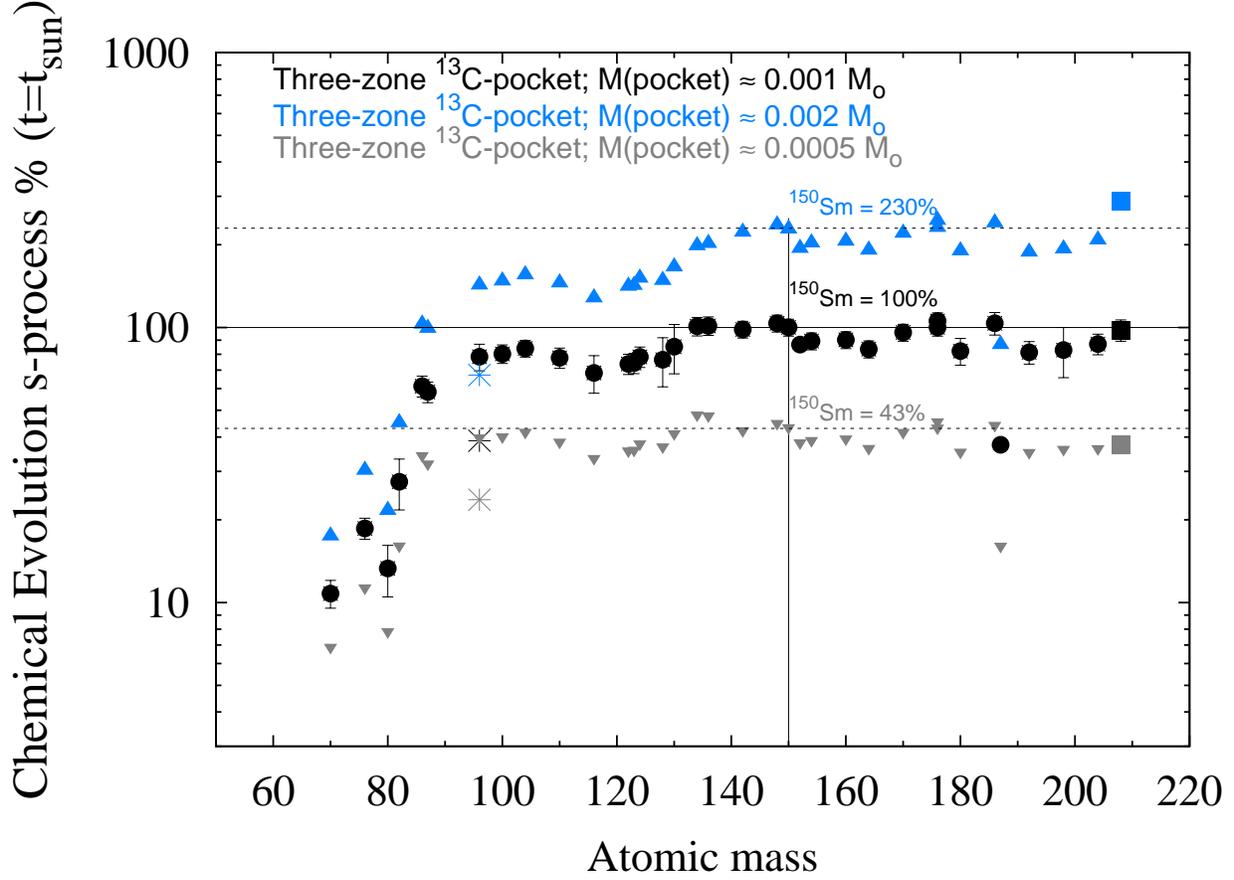}\hspace{2pc}
\caption{\label{Fig4a} The $s$-only isotopes obtained with our standard three-zone
$^{13}$C-pocket choice are represented by $filled$ $circles$; the tests with unchanged $^{13}$C 
profile, but the mass of the pocket multiplied and divided by a factor of two 
are displayed with $filled$ $triangles$ and $down$-$rotated$-$triangles$ (see text).
 GCE predictions obtained with $M$(pocket) $\sim$ 0.002 $M_\odot$ are
on average about a factor of 2 higher than solar (e.g., $^{150}$Sm = 228\%).
On the other hand, solar $s$-abundances are halved with $M$(pocket) $\sim$ 0.0005 $M_\odot$ 
(e.g., $^{150}$Sm = 43\%).
These discrepancies could be solved 
once we change the weight of the different $^{13}$C-pockets (see text). 
{\it See the electronic paper for a color version of this Figure.} }
\end{figure*}

\begin{figure*} 
\includegraphics[angle=-90,width=40pc]{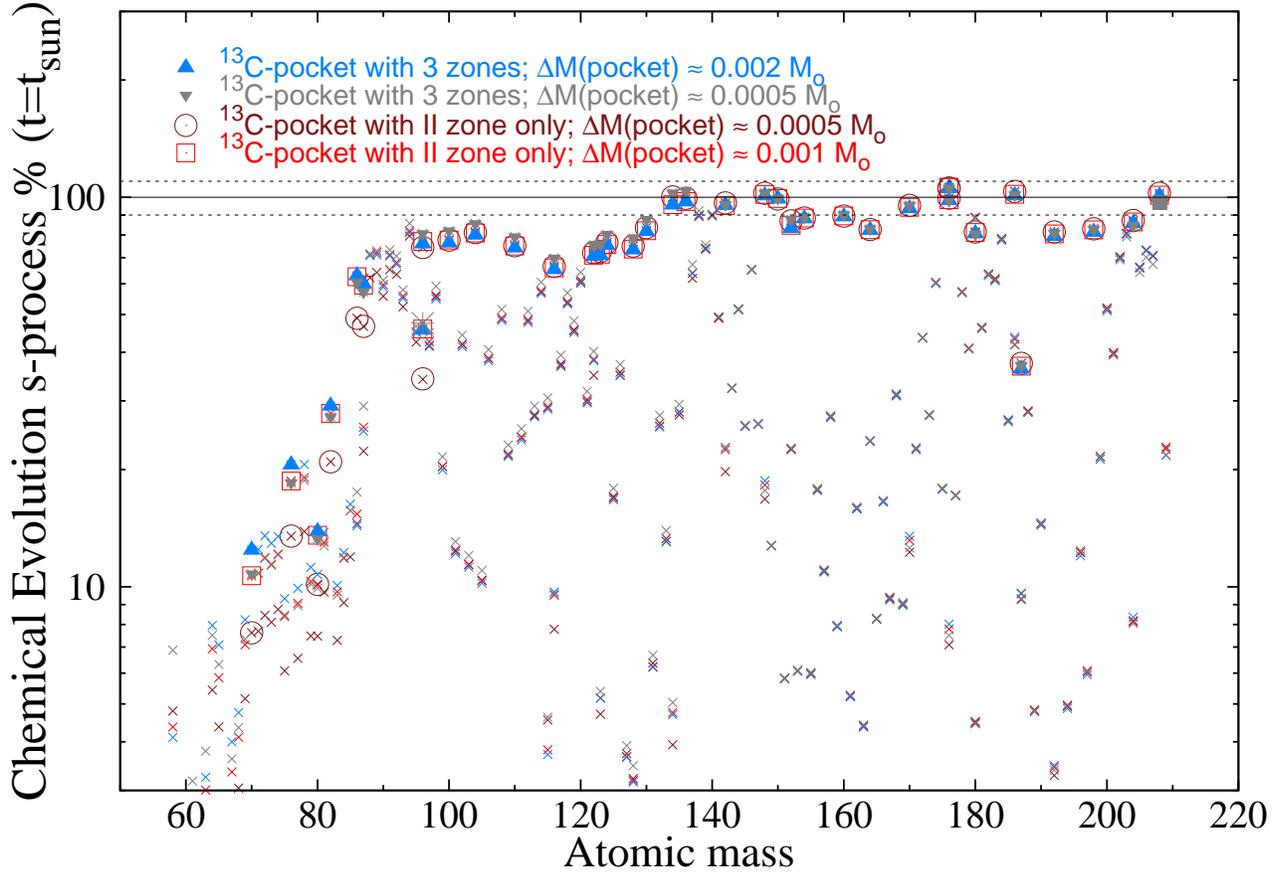}\hspace{2pc}
\caption{\label{Fig3} Effect of the $^{13}$C-pocket uncertainties on solar $s$-process 
predictions computed with GCE model. The $s$-only isotopes obtained by several tests have 
been displayed with different symbols: the tests with unchanged $^{13}$C 
profile, but the mass of the pocket multiplied and divided by a factor of two ($M$(pocket) $\sim$ 
0.002 and 0.0005 $M_\odot$) are displayed with $filled$ $triangles$ and $down$-$rotated$-$triangles$. 
We provide two additional tests under the hypothesis that only the flat profile of zone II of our 
standard $^{13}$C-pocket forms, and by assuming various mass size: $M$(pocket) 
$\sim$ 0.0005 $M_\odot$ ($empty$ $circles$) and $M$(pocket) $\sim$ 0.001 $M_\odot$ 
($empty$ $squares$), which corresponds to the mass of zone II of our standard pocket 
multiplied by a factor of two.
Results have to be compared with those obtained with our standard 
$^{13}$C-pocket choice in Fig.~\ref{Fig1}.
Note that, a different weighted range of $^{13}$C-pocket strengths 
must be adopted in order to
reproduce 100\% of solar $^{150}$Sm when changing the mass of the $^{13}$C-pocket (see text).
{\it See the electronic paper for a color version of this Figure.} }
\end{figure*}

As a first test, we increase the mass of the pocket by a 
factor of two, keeping unchanged the amount of $^{13}$C and 
$^{14}$N in the three zones of the $^{13}$C-pocket (see Fig.~\ref{Fig4a}, $filled$ $triangles$). 
For comparison we also represent with $filled$ $circles$ the $s$-process 
distribution shown in Fig.~\ref{Fig1}. 
Intuitively, if we adopt the same weighted average of the various $^{13}$C-pocket 
strengths as performed before, 
one should obtain an 
overestimation by a factor 
of two of the whole $s$-process abundance distribution. 
Effectively, we predict 200\% of solar $^{136}$Ba, 230\% of solar $^{150}$Sm and 
300\% of solar $^{208}$Pb. 
We already recalled that, in order to reconcile GCE predictions with solar abundances, we have the freedom
to change the weighted average among the various $^{13}$C-pocket strengths: specifically, we exclude 
case ST$\times$1.3 from the $^{13}$C-pocket average and we reduce the weight of case ST to 20\%
(see Fig.~\ref{Fig3}, $filled$ $triangles$).
This may be justified in the framework of the observed $s$-process spread, which, within the 
large degree of current AGB model uncertainties, we compute with a range of $^{13}$C-pocket
strengths.

As a second test, we reduce the mass of the three-zone $^{13}$C-pocket by a factor
of two (see Fig.~\ref{Fig4a}, $filled$ $down$-$rotated$-$triangles$). 
The predicted solar $s$-contributions are reduced to 48\% of solar $^{136}$Ba, 
43\% of solar $^{150}$Sm and 37\% of solar $^{208}$Pb.
As shown in Fig.~\ref{Fig3} ($filled$ $down$-$rotated$-$triangles$), this underestimation 
can be solved by fully including case ST$\times$1.3 (which in our standard GCE calculation 
is considered at $\sim$25\%, Section~\ref{results}) in the $^{13}$C-pocket average.
\\
As consequence, an increased or decreased mass of the $^{13}$C-pocket by a factor 
of two in all LMS models marginally affects the predicted solar $s$-process distribution 
obtained with a weighted average of $^{13}$C-pocket strengths.

Similar solutions are achieved by the two additional tests computed with flat $^{13}$C 
profiles (see Fig.~\ref{Fig3}): zone-II $^{13}$C-pockets with $M$(pocket) $\sim$ 0.0005 
$M_\odot$ ($empty$ $circles$),
and $M$(pocket) $\sim$ 0.001 $M_\odot$ ($empty$ $squares$), which corresponds to the mass 
of zone II of our standard pocket multiplied by a factor of two.
By giving different weights to cases ST$\times$1.3 and ST, 
which 
dominate the average of $^{13}$C-pocket strengths, the solar $s$-distribution
predicted by GCE model shows variations within the solar uncertainties:
$\la$5\% for $s$-only isotopes from $A$ = 140 to 210 and $A$ = 100 to 125, 
and up to $\sim$5\% for $^{134,136}$Ba and $^{128,130}$Xe.
More evident variations ($\ga$10\%) are displayed by few isotopes lighter
than $A$ = 100, e.g., $^{86,87}$Sr and $^{96}$Zr, which, however, receive an additional
contribution by other astrophysical sources (e.g., weak-$s$ process, $p$-process).

The deficit of the predicted solar $s$-abundances for isotopes from A = 90 to 130 remains
unchanged with
the tests shown in Fig.~\ref{Fig3}.

In summary, according to an $s$-distribution referred to $^{150}$Sm, 
the need of an additional
$s$-process between A = 90 and 130 (LEPP-s, given that $s$-only isotopes also show the
missing contribution) is confirmed by updated GCE results.
Specifically, the need of LEPP-s is independent of the most significant $^{13}$C-pocket 
tests we carried out in this paper.
\\
This result supports the presence of an $s$-process in massive stars during
the pre-explosive phases that follow the core He-burning and convective shell C-burning phases.
According to \citet{pignatari10}, the $weak$-$s$ process may produce Sr-Y-Zr in larger amount
(up to $\sim$30\%) than previously estimated by \citet{raiteri92} ($<$10\%). However, while the 
$weak$-$s$ contribution estimated for lighter trans-iron elements (e.g., Cu) is plausibly established 
because they are weakly affected by uncertainties of nuclear rates, C shell evolution, neutron 
density history, or initial composition, heavier isotopes as Sr show larger sensitivity
to stellar models and nuclear uncertainties.
Recent investigations suggest that the $s$-contribution from massive 
stars
may extend to heavier elements, with a bulk of the 
production at Sr-Y-Zr and, in minor amount, up to Te-Xe \citep{pignatari13}.

\section{Conclusions}\label{conclusions}

We study the solar abundances of heavy $s$-isotopes at the epoch of the formation of the
solar system as the outcome of nucleosynthesis occurring in AGB stars of various masses 
and metallicities.
At present, one of the major uncertainties of AGB stellar model is the formation of the $^{13}$C-pocket.
Our aim is to investigate the impact of uncertainties concerning the internal structure 
of the $^{13}$C-pocket 
on the GCE $s$-distribution, by carrying out different tests in which we modify 
the $^{13}$C and $^{14}$N 
abundances in the pocket, and the size in mass of the pocket itself.
Thereby, we obtain that GCE $s$-process predictions at the epoch of the solar-system formation
marginally depend on the choice of the
$^{13}$C profile and on the mass of the pocket when a range of $^{13}$C-pocket strengths is adopted.
The GCE model may reproduce within the solar error bars 
the $s$-contribution to isotopes with A $>$ 130.
The missing contribution to isotopes in the range between A = 90 to 130 found by 
\citet{travaglio04} is confirmed by the present analysis: an additional $s$-process (LEPP-$s$) 
is required to account for the missing component of ten $s$-only isotopes ($^{96}$Mo, $^{100}$Ru, 
$^{104}$Pd, $^{110}$Cd, $^{116}$Sn, $^{122,123,124}$Te and $^{128,130}$Xe).
Based on the tests made in this paper, the LEPP-$s$ contribution remains essentially the same,
independently of the internal structure of the $^{13}$C-pocket.
First indications in favor of this process have been analyzed and discussed by 
\citet{pignatari13}. 
\\
An additional primary contribution is being explored to account for a complementary 
light-$r$ contributions.  
In spite of promising theoretical improvements related to the explosive phases 
of massive stars and core collapse Supernovae (\citealt{arcones13} and references 
therein), as well as recent spectroscopic investigations \citep{roederer12,hansen12,hansen13}, 
a full understanding of the origin of the neutron capture elements 
from Sr up to Xe is still lacking.


\acknowledgments

We are extremely indebted with the anonymous referee for helpful comments, which have significantly
improved the structure and clarity of the paper.
S.B. was supported by a JINA Fellowship (ND Fund \#201387 and 305387). 
Numerical calculations have been sustained by B2FH Association (\url{http://www.b2fh.org/}).

\clearpage

\clearpage

\clearpage

\begin{deluxetable}{cccccccc}
\tabletypesize{\scriptsize}
\tablewidth{0pt}
\tablecaption{\label{Tab1} Solar $s$-process contributions (in \%) for isotopes from Kr to Bi
obtained with GCE model. 
GCE results by Travaglio et al. (2004) ([T04]; normalised to solar meteoritic abundances by
Anders \& Grevesse 1989), are compared with updated GCE predictions
([This work]; normalised to solar meteoritic abundances by \citealt{lodders09}). 
In columns~3 and~6 we report the percentages of the $s$-process solar contributions from 
LMS only; in columns~4 and~7 IMS $s-$process predictions are listed.
The total s process contributions from AGB stars are reported in columns~5 and~8.
Uncertainties refer to solar abundances by Anders \& Grevesse (1989; column~5) and
Lodders et al. (2009; column~8). 
In column~2, updated $main-s$ contributions from \citet{bisterzo11} are also listed
for comparison. }
\tablecolumns{10}
\tablehead{
\colhead{ }                       &
\colhead{ Main-s\tablenotemark{(1)} }                       &
\multicolumn{3}{l}{Solar $s$-process distribution [T04]} &
   \multicolumn{3}{l}{Solar $s$-process distribution [This work]} \\ 
\colhead{Isotope}                  &
\colhead{ } &  
\colhead{LMS }               &
\colhead{IMS }               &         
\colhead{(3)+(4) }               &
\colhead{LMS }               &
\colhead{IMS }               &  
\colhead{(6)+(7) }            \\
\colhead{(1)}                  &
\colhead{(2)}                  &
\colhead{(3)}                  &
\colhead{(4)}                  &
\colhead{(5)}                  &
\colhead{(6)}                  &
\colhead{(7)}                  &
\colhead{(8)}                  \\
}       
\startdata  
     $^{80}$Kr                    &   7.9                      &  13.56    &    4.06    &    17.6                 &                 12.77   &    0.52     &   13.3                      \\
     $^{82}$Kr                    &   23.8                     &  36.93    &    15.15   &    52.1                 &                 25.64   &    1.79     &   27.4                      \\
     $^{83}$Kr                    &   9.0                      &  12.28    &    4.85    &    17.1                 &                 8.92    &    0.62     &   9.5                      \\
     $^{84}$Kr                    &   11.7                     &  12.18    &    5.35    &    17.5                 &                 10.95   &    0.77     &   11.7                      \\
     $^{86}$Kr                    &   16.2                     &  21.49    &    33.96   &    55.4                 &                 11.65   &    3.58     &   15.2                      \\
     Kr\tablenotemark{(2)}        &   13.4                     &  16.66    &    11.31   &    28.0 $\pm$ 5.0       &                 12.55   &    1.35     &   13.9  $\pm$    2.9        \\
                                  &                            &           &            &                         &                         &             &                         \\
     $^{85}$Rb                    &   18.3                     &  13.66    &    8.51    &    22.2                 &                 13.68   &    1.62     &   15.3                      \\
\enddata                         
\tablenotetext{(1)}{The $main-s$ component is reproduced as an average between AGB models of $M$ = 1.5 and
3 $M_\odot$ and half solar metallicity, as in \citet{arlandini99}. Values are taken from \citet{bisterzo11}.}
\tablenotetext{(2)}{ 
Note that massive stars may produce a consistent amount of Sr-Y-Zr, and 
in minor quantity up to Te-Xe 
(\citealt{pignatari10,pignatari13}). }
\tablenotetext{(a)}{Isotopes that receive an additional contribution from $p$-process \citep{travaglio11}.}
\tablenotetext{(b)}{Overabundances with respect to solar (in percentage).}
\end{deluxetable}                

\begin{deluxetable}{ccccc}
\tabletypesize{\scriptsize}
\tablewidth{0pt}
\tablecaption{\label{Tab2} Internal structure of the $^{13}$C-pockets adopted in
the tests displayed in Fig.~\ref{Fig3}.
First, we specify the mass and the amount of $^{13}$C and $^{14}$N in the
three-zone $^{13}$C-pocket adopted so far (case ST with zones I-II-III; $M$(pocket) = 1.09E$-$3 $M_\odot$; 
represented with $filled$ $circles$). 
In $TEST$ $A$ and $B$ we leave the $^{13}$C profile unchanged, but we
multiply and divide by a factor of two the mass of the pocket ($M$(pocket) = 2$\times$1.09E$-$3 $M_\odot$; 
displayed with $filled$ $triangles$; $M$(pocket) = 0.5$\times$1.09E$-$3 $M_\odot$; 
displayed with $filled$ $down$-$rotated$-$triangles$). 
Under the hypothesis that only the flat profile of zone II of our standard $^{13}$C-pocket forms
(zone-II model, with flat $X$($^{13}$C) and $X$($^{14}$N) abundances), we provide two additional 
tests in which we change the mass involved in the pocket: $TEST$ $C$ with $M$(pocket) 
= 5.30E$-$4 $M_\odot$ ($empty$ $circles$) and $TEST$ $D$ with $M$(pocket) = 2$\times$5.30E$-$4 $M_\odot$ 
($empty$ $squares$).} 
\tablecolumns{5}          
\tablehead{
            &  zone I    &  zone II   & zone III  & \\ 
}
\startdata  
\multicolumn{5}{c}{Three-zone model with $M$(pocket) = 1.09E$-$3 $M_\odot$} \\ 
Mass($M_\odot$)  & 5.50E$-$4    &  5.30E$-$4   &   7.50E$-$6 & \\   
$X$($^{13}$C)  & 3.20E$-$3    &  6.80E$-$3   &   1.60E$-$2 & \\ 
$X$($^{14}$N)  & 1.07E$-$4    &  2.08E$-$4   &   2.08E$-$3 & \\ 
\cline{1-5}
\multicolumn{5}{c}{$TEST$ $A$: three-zone model with $M$(pocket) = 2$\times$1.09E$-$3 $M_\odot$} \\ 
Mass($M_\odot$)  & 1.10E$-$3    &  1.06E$-$3   &   1.50E$-$5 & \\   
$X$($^{13}$C)  & 3.20E$-$3    &  6.80E$-$3   &   1.60E$-$2 & \\  
$X$($^{14}$N)  & 1.07E$-$4    &  2.08E$-$4   &   2.08E$-$3 & \\
\enddata                         
\end{deluxetable}


\begin{thebibliography}{}
\bibitem[Abia et al.(2001)]{abia01}
 Abia, C., Busso, M., Gallino, R., Dom{\'i}nguez, I., Straniero, O., 
 \& Isern, J. 2001, \apj, 1117, 1134
\bibitem[Abia et al.(2002)]{abia02}
 Abia, C., Dom{\'i}nguez, I., Gallino, R. et al. 2002, \apj, 579, 817 
 \bibitem[Allen \& Barbuy(2006)]{allen06}
 Allen, D., \& Barbuy, B. 2006, \aap, 454, 917 
\bibitem[Anders \& Grevesse(1989)]{AG89}
Anders, E., \& Grevesse, N. 1989, \gca, 
53, 197
\bibitem[Aoki et al.(2002)]{aoki02}
Aoki, W., Ryan, S. G., Norris, j. E., Beers, T. C., Ando, H., \& Tsangarides, S. 2002,
\apj, 580, 1149
\bibitem[Arcones \& Thielemann(2013)]{arcones13}
Arcones, A., \& Thielemann, F.-K. 2013,
Journal of Physics G: Nuclear and Particle Physics, 40, 013201
\bibitem[Arlandini et al.(1999)]{arlandini99} 
Arlandini, C., K{\"a}ppeler, F., Wisshak, K., Gallino, R., Lugaro, M.,
Busso, M., \& Straniero, O. 1999, \apj, 525, 886
\bibitem[Asplund(2005)]{asplund05} 
Asplund, M. 2005, \araa, 43, 481
\bibitem[Bao et al.(2000)]{bao00} Bao, Z. Y., 
Beer, H., K{\"a}ppeler F., Voss, F., \& Wisshak, K. 2000, Atomic Data \& Nuclear Data Tables, 
76, 70
\bibitem[Bennett et al.(2013)]{bennett13}
Bennett, C. L., Larson, D., Weiland, J. L. et al. 2013, \apjs, 208, 20
\bibitem[Best et al.(2001)]{best01}
Best, J., Stoll, H.,  Arlandini, C., et al. 2001, Phys.\ Rev.\ C, 64, 015801
\bibitem[Bisterzo et al.(2013)]{bisterzo13}
Bisterzo, S., Travaglio, C., Wiescher, M., Gallino, R. et al. 2013, 
Journal of Physics: Conference Series (JPCS), 
 Nuclear Physics in Astrophysics VI (NPA-VI), Lisbon, Portugal, May 19th--24th,
in press (ArXiv e-prints, arXiv:1311.5381)
\bibitem[Bisterzo et al.(2011)]{bisterzo11}
Bisterzo, S., Gallino, R., Straniero, O., Cristallo, S., \& K{\"a}ppeler, F. 2011,
\mnras, 418, 284
\bibitem[Bisterzo et al.(2010)]{bisterzo10}
Bisterzo, S., Gallino, R., Straniero, O., Cristallo, S., \& K{\"a}ppeler, F. 2010,
\mnras, 404, 1529 
\bibitem[Boothroyd \& Sackmann(1988)]{boot88}
Boothroyd, A. I., \& Sackmann, I.-J. 1988, \apj, 328, 653
\bibitem[Burbidge et al.(1957)]{burbidge57}
Burbidge, E. M., Burbidge, G. R., Fowler, W. A., \& Hoyle, F. 1957, Rev.\ Mod.\ Phys., 
29, 547
\bibitem[Busso et al.(2012)]{busso12} 
Busso, M., Palmerini, S., Maiorca, E., Trippella, O., Magrini, L., \& Randich, S.
2012, 12th Symposium on Nuclei in the Cosmos, August 5 - 12,
Cairns, Australia, 20
\bibitem[Busso et al.(2007)]{busso07}
Busso, M.,  Wasserburg, G. J., Nollett, K. M., \& Calandra, A. 2007, \apj, 671 802
\bibitem[Busso et al.(2001)]{busso01} Busso, M., 
Gallino, R., Lambert, D. L., Travaglio, C., \& Smith, V. V. 2001, \apj, 557, 802
\bibitem[Cameron(1957)]{cameron57} 
Cameron, A. G. W. 1957, \aj, 62, 9
\bibitem[Clayton \& Rassbach(1967)]{clayton67} 
Clayton, D. D., \& Rassbach, M. E. 1967, \apj, 168, 69
\bibitem[Clayton \& Nittler(2004)]{clayton04} 
Clayton, D. D., \& Nittler, L. R. 2004, \araa, 42, 39 
\bibitem[Cristallo et al.(2011)]{cristallo11} 
Cristallo, S., Piersanti, L., Straniero, O., Gallino, R., Dom{\'i}nguez, I.,
Abia, C., Di Rico, G., Quintini, M., \& Bisterzo, S. 2011, \apjs, 197, 17
\bibitem[Cristallo et al.(2009)]{cristallo09} 
Cristallo, S., Straniero, O., Gallino, R., Piersanti, L., Dom{\'i}nguez, I.,
\& Lederer, M. T. 2009, \apj, 696, 797
\bibitem[Denissenkov \& Tout(2003)]{denissenkov03} 
Denissenkov, P. A., \& Tout, C. A. 2003, \mnras, 340, 722
\bibitem[Doherty et al.(2014)]{doherty14} 
Doherty, C. L., Gil-Pons, P., Lau, H. H. B., Lattanzio, J. C., \& Siess, L.
2014, \mnras, 437, 195
\bibitem[Dom{\'i}nguez et al.(1999)]{dominguez99} 
Dom{\'i}nguez, I., Chieffi, A., Limongi, M., \& Straniero, O. 1999, \apj, 524, 226 
\bibitem[D'Orazi et al.(2013)]{d'orazi13} 
D'Orazi, V.,; Lugaro, M., Campbell, S. W., Bragaglia, A., Carretta, E., Gratton, R. G.,
Lucatello, S., \& D'Antona, F. 2013, \apj, 776, 59
\bibitem[Frischknecht, Hirschi, \& Thielemann(2012)]{fris12} 
Frischknecht, U., Hirschi, R., \& Thielemann, F.-K. 2012, \aap, 538, L2
\bibitem[Gallino et al.(1998)]{gallino98} Gallino, 
R., Arlandini, C., Busso, M., Lugaro, M., Travaglio, C., Straniero, O., Chieffi, 
A., \& Limongi, M. 1998, \apj, 497, 388
\bibitem[Garc{\'i}a-Hern{\'a}ndez et al.(2013)]{garcia13}
Garc{\'i}a-Hern{\'a}ndez,  D. A., Zamora, O., Yag{\"u}e, A., Uttenthaler, S.,
Karakas, A. I., Lugaro, M., Ventura, P., \& Lambert, D. L. 2013, \aap, 555, L3
\bibitem[Garc{\'i}a-Hern{\'a}ndez et al.(2009)]{garcia09}
Garc{\'i}a-Hern{\'a}ndez, D. A., Manchado, A., Lambert, D. L., Plez, B., Garc{\'i}a-Lario, P., 
D'Antona, F., Lugaro, M., Karakas, A. I., \& van Raai, M. A. 2009, \apj, 705, 31
\bibitem[Goriely \& Siess(2004)]{goriely04}
Goriely, S., \& Siess, L. 2004, \aap, 421, L25
\bibitem[Goriely \& Mowlavi(2000)]{goriely00}
Goriely, S., \& Mowlavi, N. 2000, \aap, 362, 599 
\bibitem[Hansen et al.(2012)]{hansen12}
Hansen, C. J., Primas, F., Hartman, H., et al. 2012, \aap, 545, 31
\bibitem[Hansen et al.(2013)]{hansen13}
Hansen, C. J., Bergemann, M., Cescutti, G., Fran\c{c}ois, P., Arcones, A., 
Karakas, A. I., Lind, K., \& Chiappini, C. 2013, \aap, 551, A57 
\bibitem[Hayakawa et al.(2008)]{hayakawa08}
Hayakawa, T., Iwamoto, M., Kajino, T., Shizuma, T., Umeda, H., \& Nomoto,
K. 2008, \apj, 685, 1089
\bibitem[Heil et al.(2009)]{Heil09pasa}
M. Heil, M., Juseviciute, A., K{\"a}ppeler, F., Gallino, R., \& Pignatari, M., 
Uberseder, E., 2009, \pasa, 26, 243 
\bibitem[Heil et al.(2008)]{heil08} Heil, M.,
Winckler, N., Dababneh, S., et al. 2008, \apj, 673, 434 
\bibitem[Herwig(2005)]{herwig05}
Herwig, F. 2005, \araa, 43, 435
\bibitem[Herwig(2004)]{herwig04}
Herwig, F. 2004, \apj, 605, 425
\bibitem[Herwig, Langer \& Lugaro(2003)]{herwig03} Herwig F.
Langer N., \& Lugaro M., 2003, \apj, 593, 1056
\bibitem[Herwig et al.(1997)]{herwig97} Herwig, F., 
Bl{\"o}cker, T., Sch{\"o}nberner, D., \& El Eid, M. 1997, \aap, 324, L81
\bibitem[Howard, Meyer, \& Woosley(1991)]{howard91}
Howard, W. M., Meyer, B. S., \& Woosley, S. E. 1991, \apjl, 373, 5
\bibitem[Iben \& Renzini(1983)]{iben83}
 Iben, I. Jr, \& Renzini, A. 1983, \araa, 21, 271
\bibitem[Israelian et al.(2001)]{israelian01}
Israelian, G., Rebolo, R., Garc{\'i}a L{\'o}pez, R. J., Bonifacio, P., Molaro, P.,
Basri, G., \& Shchukina, N. 2001, \apj, 551, 833
\bibitem[Jaeger et al.(2001)]{jaeger01}
Jaeger, M., Kunz, R., Mayer, A., Hammer, J. W., Staudt, G., Kratz, K. L., \& Pfeiffer, 
B. 2001, Phys.\ Rev.\ Lett.,  
87, 202501
\bibitem[Jonsell et al.(2006)]{jonsell06}
Jonsell, K., Barklem, P. S., Gustafsson, B., Christlieb, N., Hill, V., Beers, T. C.,
Holmberg, J. 2006, \aap, 451, 651
\bibitem[K{\"a}ppeler et al.(2011)]{kaeppeler11} 
K{\"a}ppeler, F., Gallino, R., Bisterzo, S., \& Aoki, W. 2011, Rev.\ Mod.\ Phys., 
83, 157
\bibitem[K{\"a}ppeler et al.(1994)]{kaeppeler94} 
K{\"a}ppeler, F., Wiescher, M., Giesen, U., et al.
 1994, \apj, 437, 396
 \bibitem[Karakas, Garc{\'i}a-Hern{\'a}ndez, \& Lugaro(2012)]{karakas12}
Karakas, A. I., Garc{\'i}a-Hern{\'a}ndez, D. A., \& Lugaro, M. 2012, \apj, 751, 8
\bibitem[Karakas et al.(2010)]{karakas10}
 Karakas, A. I. 2010, \mnras, 403, 1413
\bibitem[Karakas \& Lattanzio(2003)]{karakas03} 
Karakas, A., \& Lattanzio, J. 2003, \pasa, 20, 279
\bibitem[Koehler \& Guber(2013)]{koehler13}
Koehler, P. E., \& Guber, K. H. 2013, Phys.\ Rev.\ C., 88, 035802
\bibitem[Lai et al.(2008)]{lai08}
Lai, D. K., Bolte, M., Johnson, J. A., Lucatello, S., 
Heger, A., \& Woosley, S. E. 2008, \apj, 681, 1524
\bibitem[Lambert et al.(1995)]{lambert95} 
Lambert, D. L., Smith, V. V., 
Busso, M., Gallino, R., \& Straniero, O. 1995, \apj, 450, 302
\bibitem[Langer et al.(1999)]{langer99} Langer, N., 
Heger, A. Wellstein, S., \& Herwig, F. 1999, \aap, 346, L37
\bibitem[Lau, Stancliffe, \& Tout(2009)]{lau09} 
Lau, H. H. B., Stancliffe, R. J., \& Tout, C. A. 2009, \mnras, 396, 1046
\bibitem[Lederer et al.(2013)]{lederer13} 
Lederer, C., Massimi, C., Altstadt, S., et al. 2013, Phys.\ Rev.\ Lett., 110, 022501
\bibitem[Lodders, Palme \& Gail(2009)]{lodders09} 
Lodders, K., Palme, H., \& Gail, H.-P. 2009, Landolt-B{\"o}rnstein - Group VI 
Astronomy and Astrophysics Numerical Data and Functional Relationships in Science 
and Technology, Edited by J.E. Tr{\"u}mper, 4B: solar system, 4.4 
\bibitem[Longland, Iliadis \& Karakas(2012)]{longland12} 
Longland, R., Iliadis, C., \& Karakas, A. I. 2012, Phys.\ Rev.\ C, 85, 065809
\bibitem[Lugaro et al.(2014)]{lugaro14}
Lugaro, M., Tagliente, G., Karakas, A. I., Milazzo, P. M., Kaeppeler, F., Davis, A. M.,
\& Savina, M. R. 2014, \apj, 780, 95 
\bibitem[Lugaro et al.(2012)]{lugaro12}
Lugaro, M., Karakas, A., Stancliffe, R. J., \& Rijs, C. 2012, \apj, 747, 2
\bibitem[Lugaro et al.(2003)]{lugaro03}
Lugaro, M., Davis, A. M., Gallino, R., Pellin, M. J., Straniero, O. \& K{\"a}ppeler, 
F. 2003, \apj, 593, 486
\bibitem[Marganiec et al.(2010)]{marganiec10}
Marganiec, J., Dillmann, I., Domingo Pardo, C., K{\"a}ppeler, F., \& Walter, S., 2010,
Phys.\ Rev.\ C, 82, 035806
\bibitem[Massimi et al.(2012)]{massimi12}
Massimi, C., Koehler, P., Bisterzo, S., et al. 2012, Phys.\ Rev.\ C, 85, 044615
\bibitem[Mel{\'e}ndez et al.(2006)]{melendez06}
Mel{\'e}ndez, J., Shchukina, N. G., Vasiljeva, I. E., \& Ram{\'i}rez, I.
2006, \apj, 642, 1082 
\bibitem[Mishenina et al.(2013)]{mishenina13}
Mishenina, T. V., Pignatari, M., Korotin, S. A., Soubiran, C., Charbonnel, C., 
Thielemann, F.-K., Gorbaneva, T. I., \& Basak, N. Yu. 2013, \aap, 552, 128 
\bibitem[Mosconi et al.(2010)]{mosconi10}
Mosconi, M., Fujii, K., Mengoni, A., et al. 2010, Phys.\ Rev.\ C, 82, 015802 
\bibitem[Mutti et al.(2005)]{mutti05}
Mutti, P., Beer, H., Brusegan, A., Corvi, F., \& Gallino, R. 2005,
AIP Conf. Proc., 769, 1327, INTERNATIONAL CONFERENCE ON NUCLEAR DATA 
FOR SCIENCE AND TECHNOLOGY, 26 September -- 1 October 2004,
Santa Fe, New Mexico (USA)
\bibitem[Nissen et al.(2002)]{nissen02}
Nissen, P. E., Primas, F. Asplund, M. \& Lambert, D. L. 2002, \aap, 390, 235
\bibitem[Palme \& Beer(1993)]{palme93}
Palme, H., \& Beer, H. 1993, in Landolt B{\"o}rnstein Group VI, Astronomy
and Astrophysics, Vol. 2A, ed. H. H. Voigt (Berlin: Springer), 196
\bibitem[Patronis et al.(2004)]{patronis04}
Patronis, N., Dababneh, S., Assimakopoulos, P. A., et al. 2004, Phys. Rev. C,
69, 025803
\bibitem[P{\'e}quignot \& Baluteau(1994)]{pequignot94}
P{\'e}quignot, D., \& Baluteau, J.-P. 1994, \aap, 283, 593
\bibitem[Piersanti, Cristallo, \& Straniero(2013)]{piersanti13}
Piersanti, L., Cristallo, S., \& Straniero, O. 2013, \apj, arXiv:1307.2017
\bibitem[Pignatari et al.(2013)]{pignatari13}
Pignatari, M., Hirschi, R., Wiescher, M. et al. 2013, \apj, 762, 31
\bibitem[Pignatari et al.(2010)]{pignatari10}
Pignatari, M., Gallino, R., Heil, M., Wiescher, M., K{\"a}ppeler, F., Herwig, F.,
\& Bisterzo, S. 2010, \apj, 710, 1557
\bibitem[Raiteri, Gallino, \& Busso(1992)]{raiteri92}
Raiteri, C. M., Gallino, R., \& Busso, M. 1992, \apj, 387, 263
\bibitem[Ram{\'i}rez, Allende Prieto, \& Lambert(2013)]{ramirez13}
Ram{\'i}rez, I., Allende Prieto, C., \& Lambert, D. L. 2013, \apj, 764, 78
\bibitem[Ram{\'i}rez, Mel{\'e}ndez, \& Chanam{\'e}(2012)]{ramirez12}
Ram{\'i}rez, I., Mel{\'e}ndez, J., \& Chanam{\'e}, J. 2012, \apj, 757, 164 
\bibitem[Rauscher(2012)]{rauscher12}
Rauscher, T. 2012, \apjl, 755, 10
\bibitem[Rauscher et al.(2002)]{rauscher02}
Rauscher, T., Heger, A., Hoffman, R. D., \& Woosley, S. E. 2002, \apj, 576, 323
\bibitem[Reifarth et al.(2012)]{reifarth12}
Reifarth, R., Dababneh, S., Heil, M., K{\"a}ppeler, F., Plag, R., Sonnabend, K.,
\& Uberseder, E. 2012, Phys. Rev. C, 85, 035802 
\bibitem[Reifarth et al.(2003)]{reifarth03}
Reifarth, R., Arlandini, C., Heil, M., et al. 2003, \apj, 582, 1251
\bibitem[Reifarth et al.(2002)]{reifarth02}
Reifarth, R., Heil, M., K{\"a}ppeler, et al. 2002, Phys.\ Rev.\ C, 66, 064603
\bibitem[Reyniers et al.(2007)]{reyniers07}
Reyniers, M., Abia, C., Van Winckel, H., Lloyd Evans, T., Decin, L.,
Eriksson, K., \& Pollard, K. 2007, \aap, 461, 641
\bibitem[Roederer(2012)]{roederer12}
Roederer, I. U. 2012, \apj, 756, 36
\bibitem[Seeger, Fowler, \& Clayton(1965)]{seeger65}
Seeger, P. A., Fowler, W. A., \& Clayton, D. D. 1965, \apjs, 11, 121
\bibitem[Serminato et al.(2009)]{serminato09}
Serminato, A., Gallino, R., Travaglio, C., Bisterzo, S., \& Straniero, O. 2009
\pasa, 26, 153
\bibitem[Sharpee et al.(2007)]{sharpee07}
Sharpee, B., Zhang, Y., Williams, R., Pellegrini, E., Cavagnolo, K., Baldwin, J. A.,
Phillips, M., \& Liu, X.-W. 2007, \apj, 659, 1265 
\bibitem[Spite et al.(2005)]{spite05}
Spite, M., Cayrel, R., Plez, B., et al. 2005, \aap, 430, 655
\bibitem[Sterling \& Dinerstein(2008)]{sterling08}
Sterling, N. C., \& Dinerstein, H. L. 2008, \apjs, 174, 158
\bibitem[Siess(2010)]{siess10}
Siess, L. 2010, \aap, 512, A10
\bibitem[Siess, Goriely, \& Langer(2004)]{siess04}
Siess, L., Goriely, S., \& Langer, N. 2004, \aap, 415, 1089
\bibitem[Smith \& Lambert(1990)]{smith90}
Smith, V. V., \& Lambert, D. L. 1990, \apjs, 72, 387 
\bibitem[Sneden, Cowan \& Gallino(2008)]{sneden08} 
Sneden, C., Cowan, J. J., \& Gallino, R. 2008, \araa, 46, 241
\bibitem[Sneden et al.(2003)]{sneden03} Sneden, 
C., Cowan, J. J., Lawler, J. E. et al. 2003, \apj, 591, 936
\bibitem[Spergel et al.(2003)]{spergel03} 
Spergel, D. N., Verde, L., Peiris, H. V. et al. 2003, \apjs, 148, 175
\bibitem[Straniero, Gallino \& Cristallo(2006)]{straniero06} 
Straniero, O., Gallino, R., \& Cristallo, S. 2006, 
Nucl.\ Phys.\ A, 777, 311
\bibitem[Straniero et al.(2003)]{straniero03} Straniero,
O.,  Dom{\'i}nguez, I.,  Cristallo, S., \& Gallino, R. 2003, \pasa, 20, 389
\bibitem[Straniero et al.(2000)]{straniero00}
Straniero, O., Limongi, M., Chieffi, A., Dom{\'i}nguez, I., Busso, M., \& Gallino, R.
2000, Mem.\ Soc.\ Astron.\ Ital., 71, 719
\bibitem[Straniero et al.(1995)]{straniero95} 
Straniero, O., Gallino, R., Busso, M., Chieffi, A., Raiteri, C. M., Limongi, M., 
Salaris, M. 1995, ApJ, 440, L85
\bibitem[Tagliente et al.(2011)]{tagliente11}
Tagliente, G., Milazzo, P. M., Fujii, K., et al. 2011, Phys.\ Rev.\ C, 84, 015801 
\bibitem[Tagliente et al.(2011a)]{tagliente11a}
Tagliente, G., Milazzo, P. M., Fujii, K., et al. 2011a, Phys.\ Rev.\ C, 84, 055802
\bibitem[Tagliente et al.(2010)]{tagliente10}
Tagliente, G., Milazzo, P. M., Fujii, K., et al. 2010, Phys.\ Rev.\ C, 81, 055801 
\bibitem[Thielemann et al.(2011)]{thielemann11}
Thielemann, F.-K., Arcones, A., K{\"a}ppeli, R. et al. 2011,
Progress in Particle and Nuclear Physics, 66, 346
\bibitem[Tomkin \& Lambert(1999)]{tomkin99}
Tomkin, J., \& Lambert, D. L. 1999, \apj, 523, 234
\bibitem[Toukan \& K{\"a}ppeler(1990)]{toukan90}
Toukan, K., \& K{\"a}ppeler, F. 1990, \apj, 348, 357 
\bibitem[Travaglio et al.(2011)]{travaglio11}
Travaglio, C., R{\"o}pke, F. K., Gallino, R., \& Hillebrandt, W., 2011, ApJ, 739, 93
\bibitem[Travaglio et al.(2004)]{travaglio04} Travaglio,
C., Gallino, R., Arnone, E., Cowan, J., Jordan, F., \& Sneden, C. 2004, \apj, 601, 864
\bibitem[Travaglio et al.(2004a)]{travaglio04SNII} 
Travaglio, C., Hillebrandt, W., Reinecke, M., \& Thielemann, F.-K. 2004a, \aap,
425, 1029
\bibitem[Travaglio et al.(2001)]{travaglio01} Travaglio,
C., Gallino, R., Busso, M., \& Gratton, R. 2001, \apj, 549, 346
\bibitem[Travaglio et al.(1999)]{travaglio99} Travaglio, 
C., Galli, D., Gallino, R., Busso, M., Ferrini, F., \& Straniero, O. 1999, \apj, 521, 691
\bibitem[Van Eck et al.(2001)]{vaneck01}
Van Eck, S. Goriely, S., Jorissen, A., \& Plez, B. 2001, Nature, 412, 793
\bibitem[van Raai et al.(2012)]{vanraai12}
van Raai, M. A., Lugaro, M., Karakas, A. I., Garc{\'i}a-Hern{\'a}ndez, D. A., \& Yong, D.
2012, \aap, 540, 44
\bibitem[Ventura et al.(2013)]{ventura13} 
Ventura, P., Di Criscienzo, M., Carini, R., \& D'Antona, F. 2013, \mnras, 431, 3642
\bibitem[Ventura \& D'Antona(2005)]{ventura05} 
Ventura, P., \& D'Antona, F. 2005, \aap, 431, 279
\bibitem[Ventura \& D'Antona(2005a)]{ventura05b} 
Ventura, P., \& D’Antona, F. 2005a, \aap, 439, 1075
\bibitem[Winckler et al.(2006)]{winckler06}
Winckler, N., Dababneh, S., Heil, M. et al. 2006, \apj, 647, 685
\bibitem[Wisshak et al.(2004)]{wisshak04}
Wisshak, K., Voss, F., Arlandini, C., K{\"a}ppeler, F., Heil, M., Reifarth, R.,
Krticka, M., \& Becvar, F.  2004, Phys.\ Rev.\ C,69, 055801  
\bibitem[Yong et al.(2008)]{yong08}
Yong, D., Lambert, D. L., Paulson, D. B., \& Carney, B. W. 2008, \apj, 673, 854
\bibitem[Zamora et al.(2009)]{zamora09}
Zamora, O., Abia, C., Plez, B., Domınguez, I., \& Cristallo, S. 2009, \aap, 508, 909 
\bibitem[Zinner(2007)]{zinner07}
Zinner, E. 2007, Presolar grains. Meteorites, Planets, and Comets, Treatise on Geochemistry, 
eds A.M. Davis, H.D. Holland, and K.K. Turekian (Elsevier, Oxford), published electronically 
at http://www.sciencedirect.com/science/referenceworks/9780080437514, 2nd Ed, Vol. 1. 


\end{thebibliography}
\end{document}